\documentclass{aa}  

\usepackage{float}            
\usepackage{graphicx}
\usepackage{rotating}         
\usepackage{txfonts}
\usepackage{subfigure}
\usepackage{subfig}
\usepackage{amssymb}
\usepackage{longtable}
\usepackage{pdflscape}
\usepackage{natbib}
\bibpunct{(}{)}{;}{a}{}{,} 
\usepackage{url}

\def\la{\mathrel{\mathchoice {\vcenter{\offinterlineskip\halign{\hfil
$\displaystyle##$\hfil\cr<\cr\sim\cr}}}
{\vcenter{\offinterlineskip\halign{\hfil$\textstyle##$\hfil\cr
<\cr\sim\cr}}}
{\vcenter{\offinterlineskip\halign{\hfil$\scriptstyle##$\hfil\cr
<\cr\sim\cr}}}
{\vcenter{\offinterlineskip\halign{\hfil$\scriptscriptstyle##$\hfil\cr
<\cr\sim\cr}}}}}
\def\ga{\mathrel{\mathchoice {\vcenter{\offinterlineskip\halign{\hfil
$\displaystyle##$\hfil\cr>\cr\sim\cr}}}
{\vcenter{\offinterlineskip\halign{\hfil$\textstyle##$\hfil\cr
>\cr\sim\cr}}}
{\vcenter{\offinterlineskip\halign{\hfil$\scriptstyle##$\hfil\cr
>\cr\sim\cr}}}
{\vcenter{\offinterlineskip\halign{\hfil$\scriptscriptstyle##$\hfil\cr
>\cr\sim\cr}}}}}

\newcommand{\beq}{\begin{equation}}
\newcommand{\eeq}{\end{equation}}
\newcommand{\bdi}{\begin{displaymath}}
\newcommand{\edi}{\end{displaymath}}

\begin{document} 

\titlerunning{The segregation of starless and protostellar clumps}

\title{The segregation of starless and protostellar clumps in the Hi-GAL $\ell=224^{\circ}$ region.}


\authorrunning {L.~Olmi et.~al.}  

\author{
L. Olmi \inst{\ref{inst1},\ref{inst2}} \and
M. Cunningham \inst{3} \and
D. Elia \inst{4} \and
P. Jones \inst{3} 
}

\institute{
          INAF, Osservatorio Astrofisico di Arcetri, Largo E. Fermi 5,
          I-50125 Firenze, Italy,  \email{olmi.luca@gmail.com} \label{inst1}
\and
          University of Puerto Rico, Rio Piedras Campus, Physics Dept., Box 23343,
          UPR station, San Juan, Puerto Rico, USA  \label{inst2}
\and
	  School of Physics, University of New South Wales, Sydney NSW 2052, Australia \label{inst3}
\and
	  INAF, Istituto di Astrofisica e Planetologia Spaziali, Via Fosso del Cavaliere 100, I-00133 Roma, Italy \label{inst4} 
          }

\date{Received Month Date, 2015; accepted Month Date, 2015}

\abstract
{ Stars form in dense, dusty structures, which are embedded in larger clumps of molecular clouds
often showing  a clear filamentary structure on large scales ($\ga 1$\,pc). The origin 
(e.g., turbulence or gravitational  instabilities) and evolution of these
filaments, as well as their relation to clump and core formation, are not yet fully understood.
A large sample of both starless and protostellar clumps can now be found
in the Herschel Hi-GAL (Herschel Infrared GALactic Plane Survey)
key project, which also provides striking images
of the filamentary structure of the parent molecular clouds. Recent results indicate that 
populations of clumps on and off filaments may differ.
}
{One of the best-studied regions in the Hi-GAL survey can be observed toward the $\ell=224^{\circ}$ field.
Here, a filamentary region has been studied and it has been found that protostellar clumps are mostly located
along the main filament, whereas starless clumps are detected off this filament
and are instead found on secondary, less prominent filaments. We want to investigate
this segregation effect and how it may affect the clumps properties.
}
{We mapped the $^{12}$CO$\,(1-0)$ line and its main three
isotopologues toward the two most prominent filaments observed toward the $\ell=224^{\circ}$ field using the 
Mopra radio telescope, in order to set observational constraints on the dynamics of these structures and the
associated starless and protostellar clumps. 
}
{Compared to the starless clumps, the protostellar clumps are more luminous, more turbulent and 
lie in regions where the filamentary ambient gas shows larger linewidths. 
We see evidence of gas flowing along the main filament, but we do not find any signs of 
accretion flow from the filament onto the Hi-GAL clumps. We analyze the radial column density profile of the 
filaments and their gravitational stability.
}
{The more massive and highly fragmented main filament appears to be thermally supercritical and gravitationally bound, 
assuming that all of the non-thermal motion is contributing thermal-like support,
suggesting a later stage of evolution compared to the secondary filament. The status and evolutionary phase of the
Hi-GAL clumps would then appear to correlate with that of the host filament.
}


\maketitle

\section{Introduction}
\label{sec:intro}

Stars form in dense, dusty structures, which are embedded in larger
clumps of molecular clouds. Here, the term clump
refers to any compact density enhancement 
(with size $\sim 0.1 - 1\,$pc and density $\sim 10^4 - 10^5\,$cm$^{-3}$), as identified
by source-extraction algorithms, and core (size $\la 0.01 - 0.1\,$pc and density $\ga 10^5\,$cm$^{-3}$)
refers to the clump sub-structures which are going to form stars.
A large sample of both starless (i.e., with no sign of active star formation, and not
necessarily gravitationally bound)
and protostellar clumps can now be found
in the {\it Herschel} Hi-GAL (Herschel Infrared GALactic Plane Survey)
key project (\citealp{molinari2010PASP}). The Hi-GAL survey also provides striking images
of the filamentary structure of the parent molecular clouds on large scales ($\ga 1$\,pc),
which now allow to study the physical connection between the clumps/cores and the
filaments.
Filamentary structures were known to be present in molecular clouds even before the
recent large scale star-formation surveys at far-infrared wavelengths.
The origin (e.g., turbulence or gravitational  instabilities) and evolution of these
filaments, as well as their relation to star formation, are not yet fully understood.

Observationally, filaments show complex structures with multiple branches and hubs,
where the dense clumps/cores detected in the Hi-GAL survey are more frequently found.
The observed similarity between the mass distribution of the clumps and the stellar
initial mass function (IMF) supports the accepted idea that clumps may fragment
into sub-structures (cores or fragments) which then form stars or even clusters of stars.
Therefore, in order to draw a more complete picture of star formation it is
fundamental to study the transition phase from the observed clump mass function
(CMF) to the IMF, and to understand how the presence of filaments affects the formation of
clumps and their mass distribution.

%
\begin{figure*}
\centering
\includegraphics[width=18.1cm,angle=0]{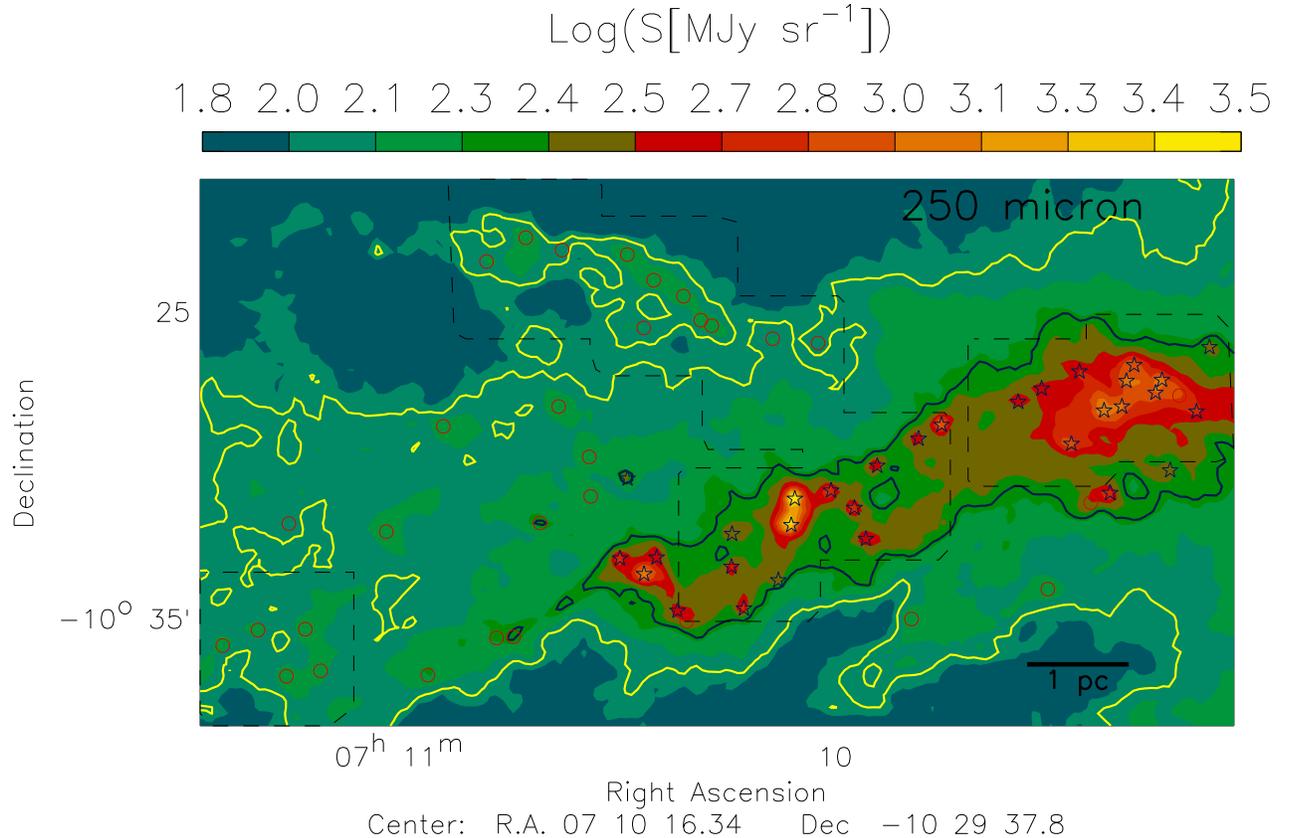}
\caption{
SPIRE $250\,\mu$m emission observed toward the $\ell=224^{\circ}$ region, with a (logarithmic) color-scale in MJy\,sr$^{-1}$
(\citealp{elia2013}). Blue star symbols represent protostellar clumps while the red empty circles represent
starless clumps. The dashed, black contour shows the region mapped with the Mopra telescope. 
The yellow and blue solid contours correspond to about 100 and 200\,MJy\,sr$^{-1}$, respectively, and 
approximately trace the secondary filament toward the north (besides to low-level emission from 
the main filament) and the main filament.
}
\label{fig:map250}
\end{figure*}

Two interesting recent results have been published as part of the
{\it Herschel} Gould Belt Survey (HGBS, \citealp{andre2010}).
\citet{polychroni2013} studied the L 1641 molecular clouds in Orion A and found
that most ($\simeq 70\,$\%) of the starless clumps were located on filaments.
They also found that the two clump populations (on and off the identified filaments) had
distinctly different CMF. In a separate work, \citet{arzoumanian2013}
investigated the gas velocity dispersions of a sample of filaments, also detected as part
of the HGBS in the IC 5146, Aquila, and Polaris interstellar clouds.
They found that these filaments could be divided into two regimes: {\it (i)} thermally subcritical
filaments, gravitationally unbound; and {\it (ii)} thermally supercritical filaments, which have
higher velocity dispersions and are self-gravitating. These authors propose that,
as they contract and accrete material from the background cloud, supercritical filaments
are expected to fragment into clumps and form (proto)stars (as supported by some
theoretical models, e.g., \citealp{pon2011}). On the other hand,
\citet{schneider2010} analyzed the dynamics of the DR21 high-mass
star-forming region and found that its filamentary morphology, and the existence of sub-filaments,
can be explained if the DR21 filament was formed by the convergence of gas flows on large scales.

In their works, \citet{polychroni2013} completely ignore the protostellar clumps,
while \citet{arzoumanian2013} and \citet{schneider2010} do not correlate the study of 
the filament dynamics with the presence of starless and protostellar clumps. In addition,
the massive DR21 region is quite different from the molecular clouds studied in the 
HGBS, and thus the question arises of whether different filament dynamics
may lead to different (low- and high-mass) star formation.
Molecular line mapping of a larger sample of filaments is thus required,
in order to set stronger observational constraints on the dynamics of these structures.

\section{ Observations}
\label{sec:obs}

\subsection{Hi-GAL observations}
\label{sec:HIGAL}

The Hi-GAL survey offers the best opportunity
to look at large clump populations in various clouds with different
physical conditions, while using self-consistent analysis to derive their physical
parameters (see, e.g., \citealp{elia2010, elia2013}, \citealp{olmi2013}).
Mass and other physical parameters of the Hi-GAL clumps were derived from
a single-temperature spectral energy distribution (SED) model applied to
suitable candidates in the Hi-GAL catalog. To extract candidate sources, a first
catalog based on image photometry was compiled in each of the observed Hi-GAL fields,
identifying the sources detected in the five different bands based on simple positional
association.

%
\begin{figure*}
\centering
\includegraphics[width=9.0cm,angle=0]{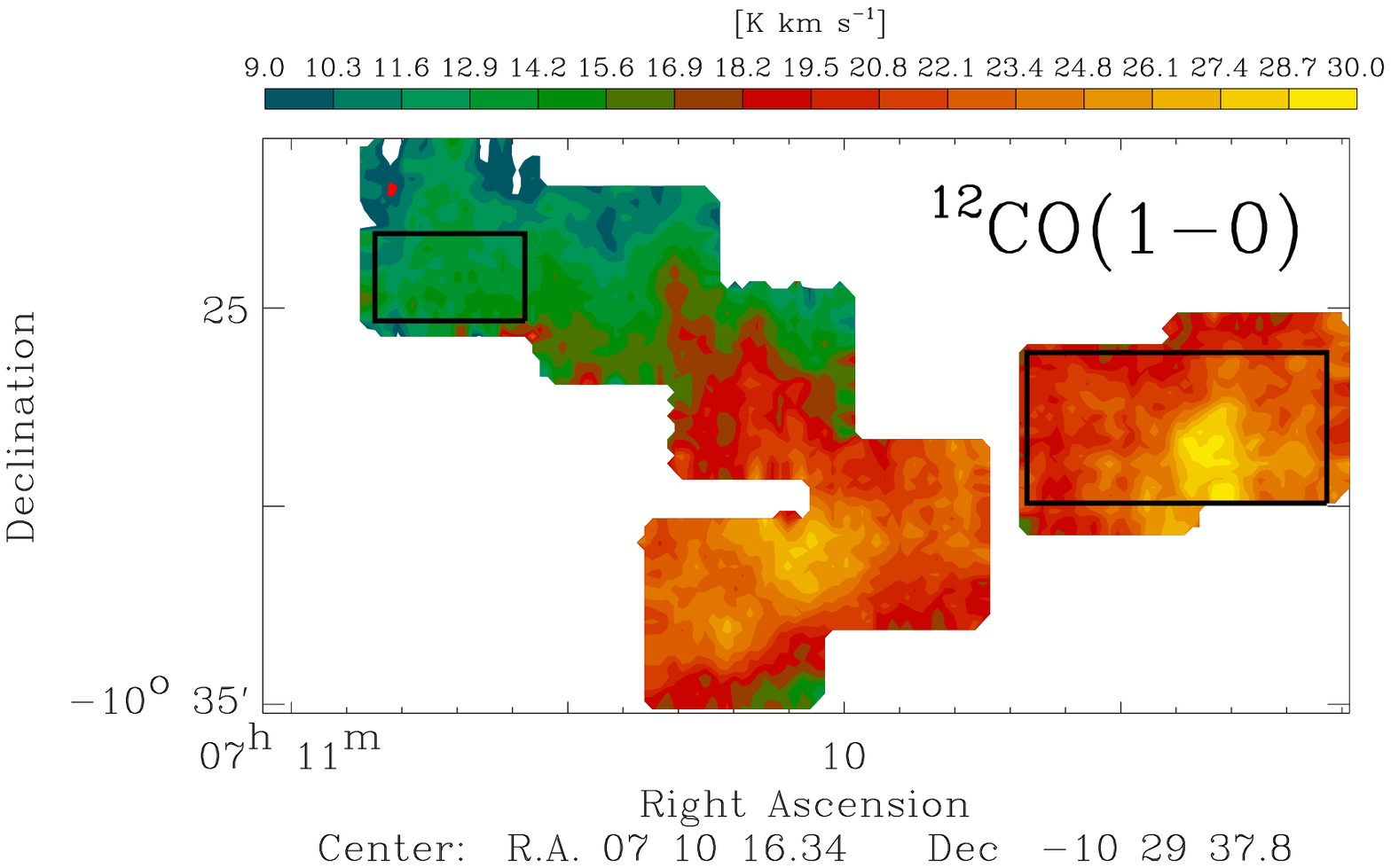}
\includegraphics[width=9.0cm,angle=0]{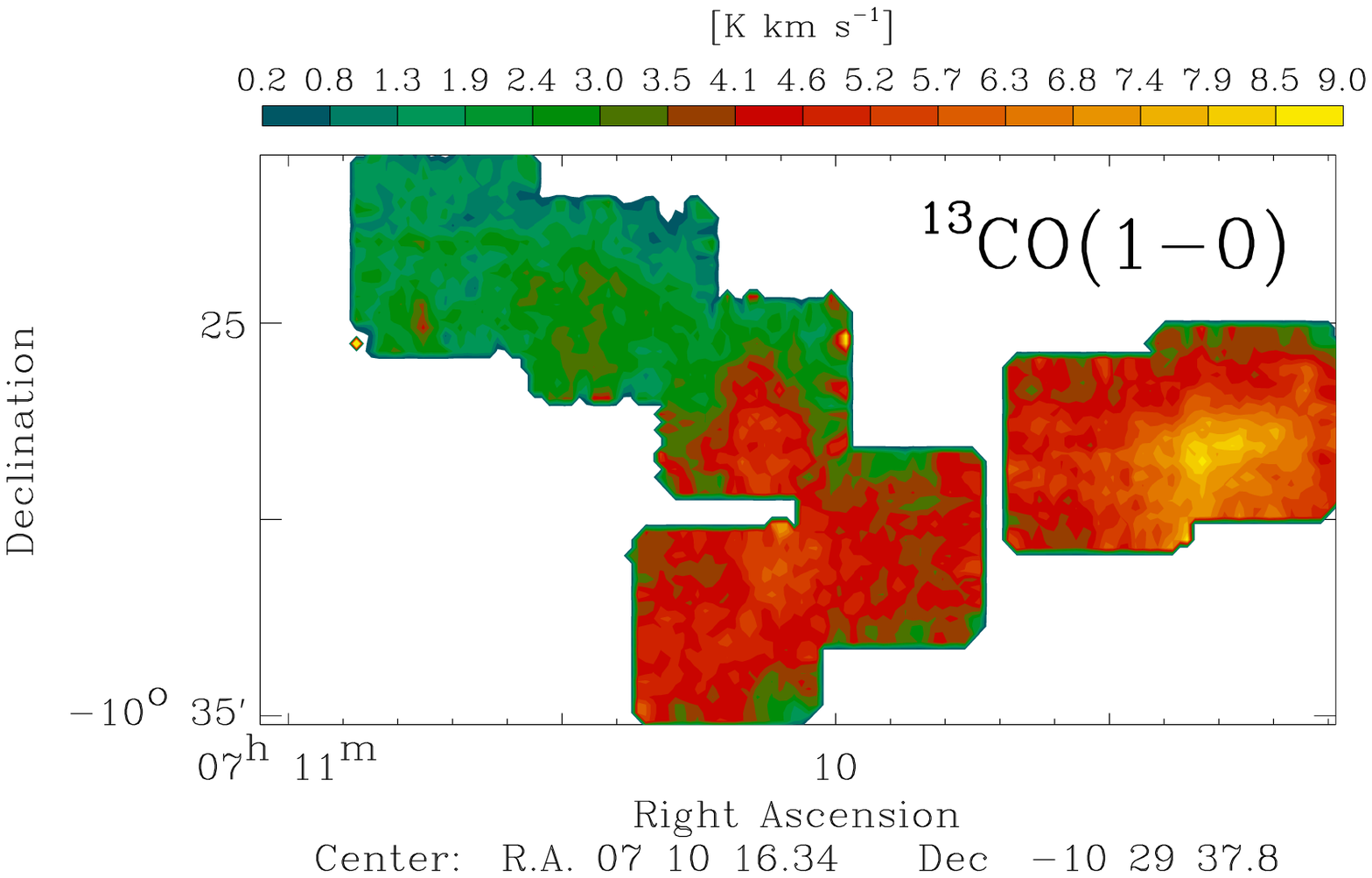}
\includegraphics[width=9.0cm,angle=0]{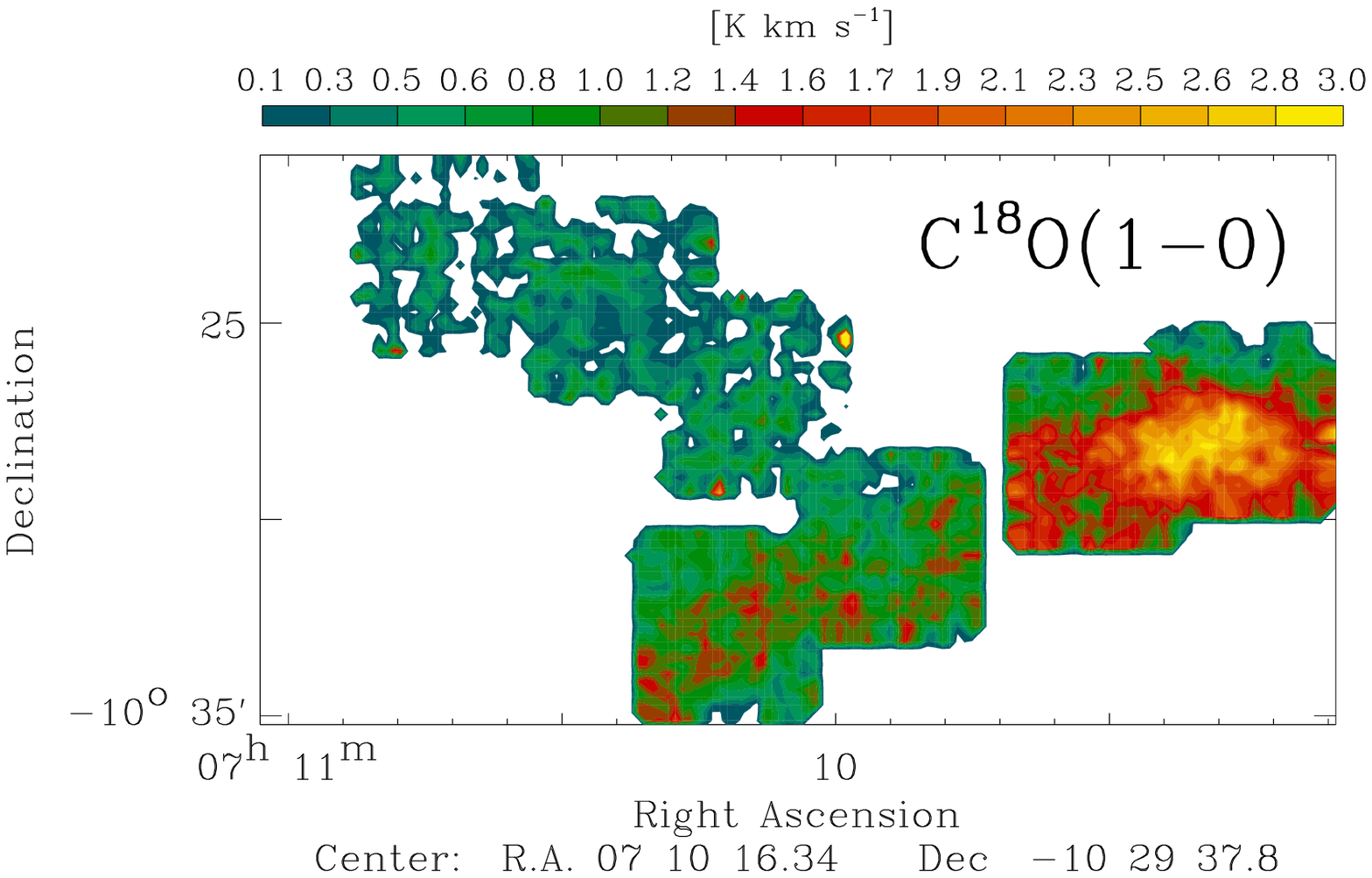}
\includegraphics[width=9.0cm,angle=0]{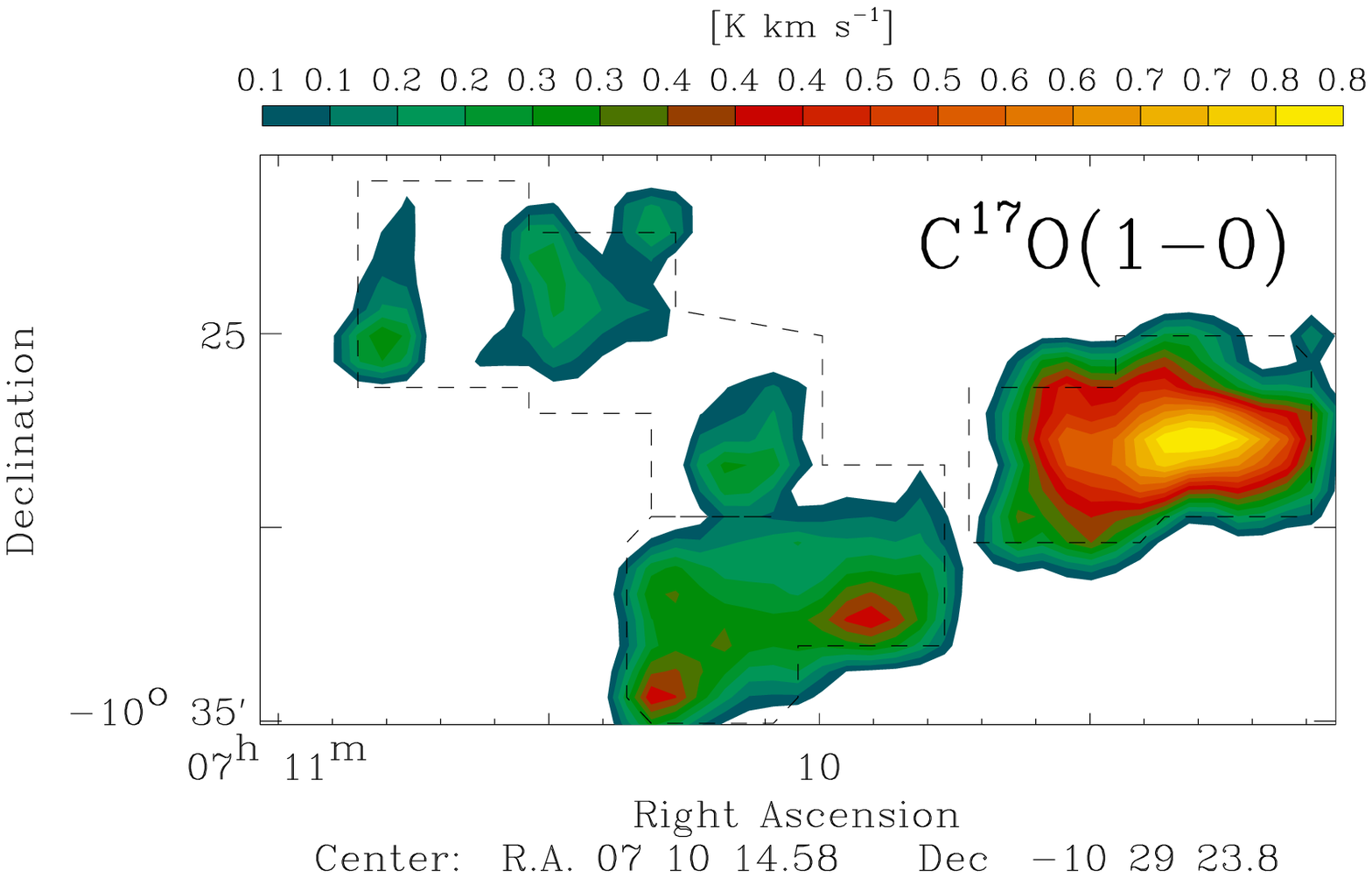}
\caption{Maps of the line integrated intensity, $\int T_{\rm A}^\star {\rm d}v$,
observed toward the $\ell=224^{\circ}$ region
in the various CO isotopologues. From left to right and from top to bottom:
$^{12}$CO$(1-0)$, $^{13}$CO$(1-0)$, C$^{18}$O$(1-0)$ and C$^{17}$O$(1-0)$.
The map of C$^{17}$O$(1-0)$ has been convolved to an 80\arcsec beam and regridded
to a 40\arcsec spacing to increase the SNR.
The black boxes represent two of the regions where spectra have been averaged and
shown in Fig.~\ref{fig:avgspectraXS}. 
}
\label{fig:maps}
\end{figure*}

One of the best-studied regions in the Hi-GAL survey is found in the $\ell= 217^{\circ}  - 224^{\circ}$ field.
The filamentary structure in this field has been studied by \citet{schisano2014}, while
\citet{elia2013} identified a well-defined and isolated filamentary cloud, 
at $\ell \simeq 224^{\circ}$ and a distance of about 1.0\,kpc, with clear signs of star formation (see Fig.~\ref{fig:map250}).
The most peculiar property of this cloud is the segregation observed
between starless and protostellar clumps. In fact, protostellar clumps are mostly found
along the main filament (filament here is used qualitatively to indicate an elongated 
or threadlike structure, and we do not use a specific operational definition; see also Sect.~\ref{sec:cdprof}), 
whereas starless clumps are detected off this filament
and are instead found on secondary, less prominent filaments (see Fig.~\ref{fig:map250}).
Therefore, also in this region most clumps are indeed located on
filaments, as already noted by \citet{polychroni2013}, but the segregation in
terms of clump type observed here has not yet been reported.

\subsection{Mopra}
\label{sec:mopra}

Our observations were carried out with the ATNF Mopra 22-m telescope\footnote{The Mopra 
radio telescope is part of the Australia Telescope National Facility (ATNF) which is 
funded by the Commonwealth of Australia for operation as a National Facility managed by CSIRO.} 
in Australia, in June 2014. 
We observed the $^{12}$CO$\,(1-0)$ (115.271202\,GHz), $^{13}$CO$\,(1-0)$ (110.201353\,GHz),
C$^{18}$O$\,(1-0)$ (109.782173\,GHz) and C$^{17}$O$\,(1-0)$ (112.359280\,GHz) spectral lines.
The $^{12}$CO$\,(1-0)$ line is known to be optically thick, and thus the other three isotopologues
were also observed to provide optical depth and line profile information.
Unfortunately, except for $^{13}$CO$(1-0)$, their intensity was in
general too weak to allow derivation of column densities throughout the map
and other useful physical and kinematical parameters, and thus most of our analysis
will be based on $^{13}$CO$(1-0)$ and partly on C$^{18}$O$(1-0)$ 
(from now on the $(1-0)$ will
be omitted for simplicity when referring to any of the observed transitions).

We used the single pixel broadband spectrometer, MOPS,
which allows for multiple spectral windows to be observed simultaneously.
Specifically, MOPS was used in its zoom mode, which allowed to split the 8.3~GHz instantaneous 
band in up to 16 zoom bands to focus on different molecules of interest. Each sub-band 
in zoom mode was 137.5 MHz wide and had 2$\times$4096 channels, and the spectral resolution 
was $\simeq 0.09\,$km\,s$^{-1}$ for the 90 GHz receiver. 
During our observations the system temperature was
typically comprised in the range $\simeq 240 - 450\,$K ($^{13}$CO and C$^{18}$O),
$\simeq 280 - 500\,$K (C$^{17}$O) and $\simeq 520 - 920\,$K ($^{12}$CO).
We reached a RMS sensitivity in $T_{\rm A}^\star$ units of
about $\sim 0.12 - 0.18 \,$K ($^{13}$CO, C$^{18}$O and C$^{17}$O) 
and $\sim 0.25 - 0.40 \,$K ($^{12}$CO)
after rebinning to a 0.20\,km\,s$^{-1}$ velocity resolution.
%
%
The beam full width at half maximum (FWHM) was about $38\,$arcsec and the
pointing was checked every hour by using a SiO maser as a reference source (\citealp{inder2013}).
Typically, pointing errors were found to be $\sim 5-10\,$arcsec.
The parameter $\eta_{\rm mb}$ to convert from antenna
temperature to main-beam brightness temperature has been assumed to be 0.44 at 100\,GHz
and 0.49 at 86\,GHz  (\citealp{ladd2005}).

The single-point observations were performed in position-switching mode, whereas
the spectral line maps, composed of individual tiles of size mostly $5\times 5\,$arcmin$^2$,
were obtained using the Mopra on-the-fly mapping mode, scanning in both right ascension and declination.
A total of 8 tiles were observed (see Figures~\ref{fig:map250} and \ref{fig:maps}): 
seven tiles are mostly contiguous, while an additional tile was observed off the main and secondary 
filaments and covers an isolated group of Hi-GAL clumps.
The data were reduced with {\sc livedata} and {\sc gridzilla} for the
bandpass correction and
gridding\footnote{\tt http://www.atnf.csiro.au/computing/software/livedata/\\index.html}.
The fits cubes were taken into {\sc miriad}\footnote{ {\tt http://bima.astro.umd.edu/miriad/}} 
for a 3-pt hanning smoothing and an additional first-order correction to the baselines.
The data were then analyzed with the {\tt xs}\footnote{ {\tt http://www.chalmers.se/rss/oso-en/observations/\\data-reduction-software}} 
package of the Onsala Space Observatory, and
they were also imported into {\sc CLASS}\footnote{CLASS is part of the
{\sc GILDAS} software package developed by IRAM.}
and IDL\footnote{\tt http://www.exelisvis.com/ProductsServices/IDL.aspx}
for line parameter measurement and further analysis.

%
%
%

%
%
\begin{figure}
\centering
\includegraphics[width=9.3cm,angle=0]{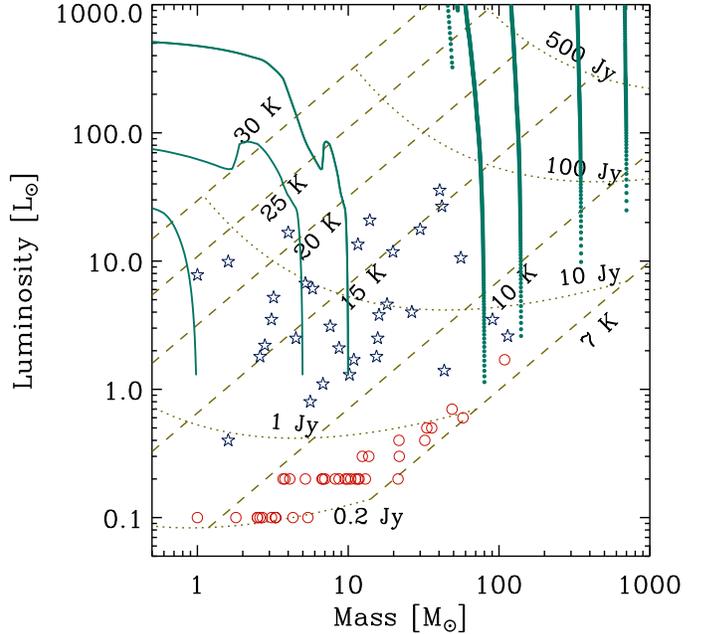}
\caption{
Clump bolometric luminosity vs. mass, as derived from the Hi-GAL observations.
Symbols for starless and protostellar sources are as in Fig.~\ref{fig:map250}.
The dotted lines represent loci of constant 250$\,\mu$m
flux density, ranging from 0.1 to 500\,Jy, assuming a modified blackbody 
spectral energy distribution with $\beta = 1.5$ and a fixed distance of $\sim 1\,$kpc.
Roughly orthogonal to these are loci (dashed lines) at constant temperature, 
for the same modified blackbody.
The green solid lines and small green filled circles represent the evolutionary tracks 
for the low- and high-mass regimes, respectively, taken from \citet{molinari2008}.
}
\label{fig:LvsM}
\end{figure}

\section{Results}
\label{sec:results}


\subsection{Physical properties of the Hi-GAL clumps}
\label{sec:clumps}


Before analysing the results of the spectral line observations we first show in Fig.~\ref{fig:LvsM} 
a plot of the bolometric luminosities versus the masses of the Hi-GAL clumps 
(see also \citealp{elia2013}).  In this plot starless and protostellar sources are 
represented by the red empty circles and blue star signs,
respectively, and one can note immediately that they populate quite different regions of the diagram.
The regions occupied by the starless and protostellar sources can be characterized by using
the evolutionary tracks discussed by \citet{molinari2008}, who proposed an evolutionary
sequence for protostars in terms of two parameters: the envelope mass and the
bolometric luminosity. These authors suggest an evolutionary sequence which is concentrated
in two main phases: protostars first accrete mass from their envelopes, and later disperse
their envelopes by winds and/or radiation. In Fig.~\ref{fig:LvsM} the calculated evolutionary
tracks thus rise upward in luminosity (almost vertically in the case of high-mass protostars)
during the accretion phase, and then proceed horizontally to the left (to lower masses) during
the envelope dispersal phase.

As already noted by  \citet{elia2013} protostellar and starless sources populate 
quite different regions of the diagram, corresponding to the accreting core phase in 
the first case, and to a quiescent or collapsing core in the second, respectively.
In fact, the protostellar sources occupy regions of the diagram at higher temperature, 
as shown by the loci at constant temperature of a modified blackbody (dashed lines).
Although this plot cannot be used to obtain a classical single-YSO classification 
(see \citealp{elia2013} for a discussion of this point), it suggests as expected 
an additional type of evolutionary segregation between
starless and proto-stellar clumps, besides to the observed positional segregation discussed in Section~\ref{sec:HIGAL}.

%
%
%
%
\begin{figure*}
\centering
\includegraphics[width=6cm,angle=0]{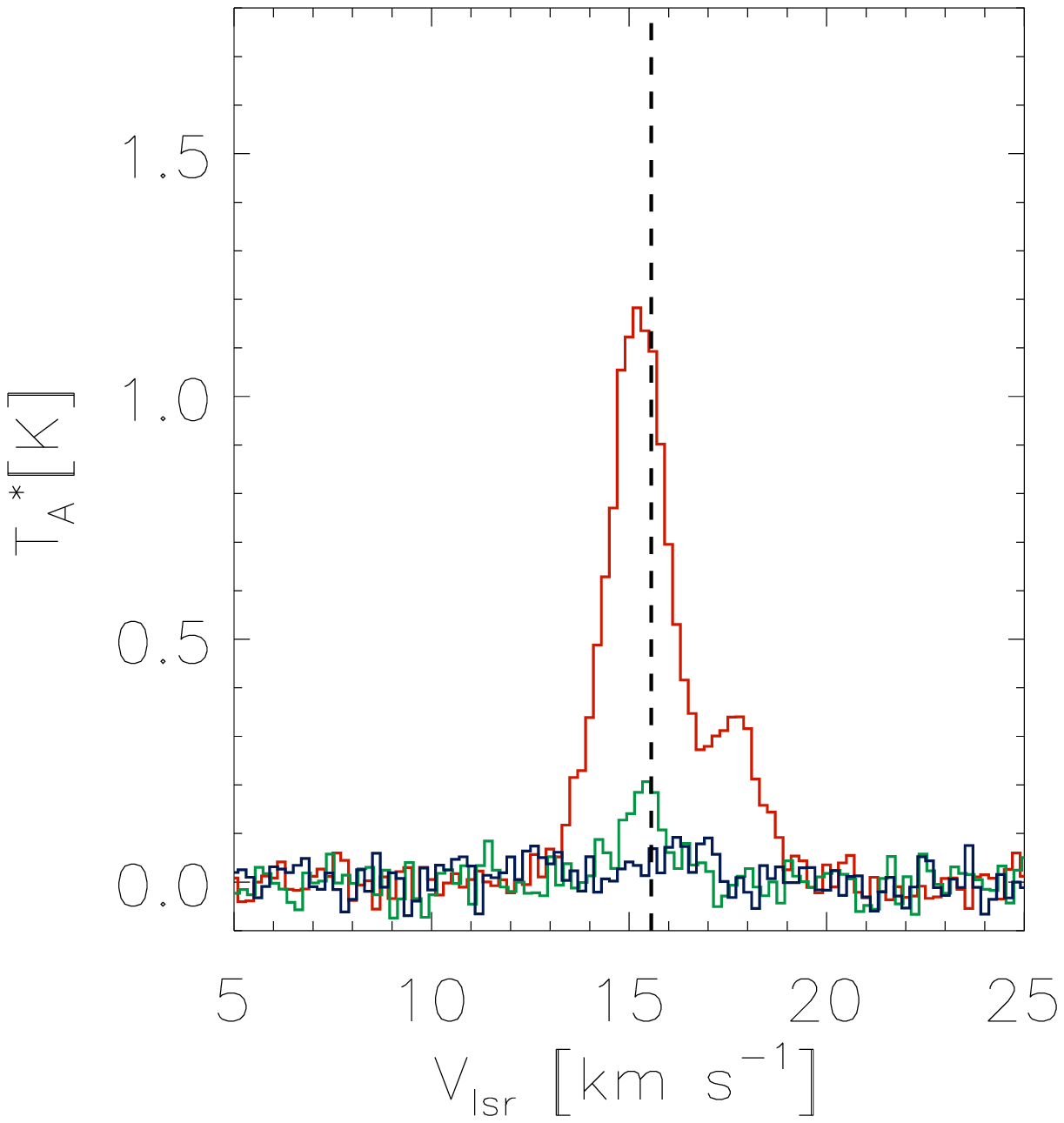}    
\includegraphics[width=6cm,angle=0]{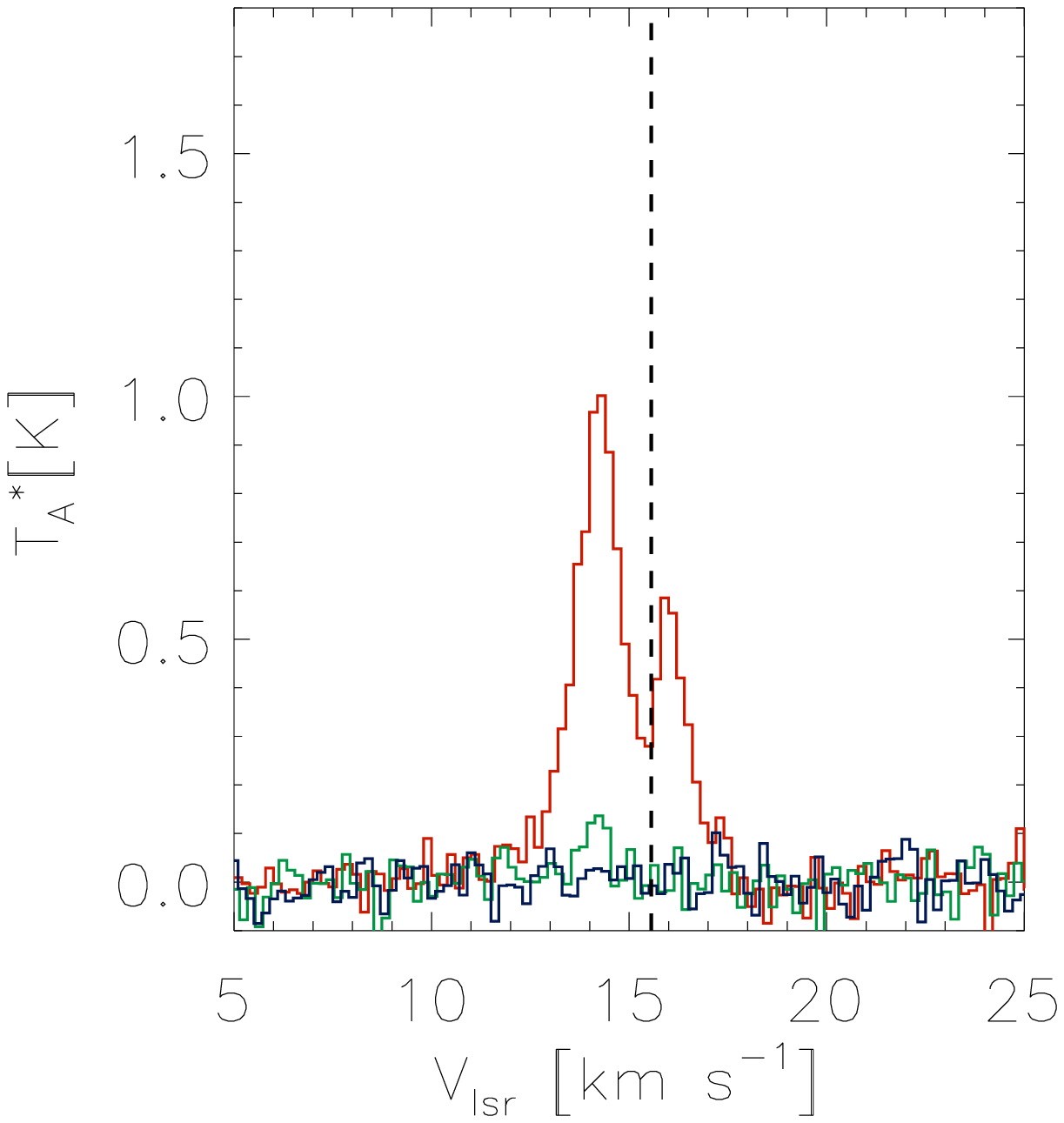}    
\includegraphics[width=6cm,angle=0]{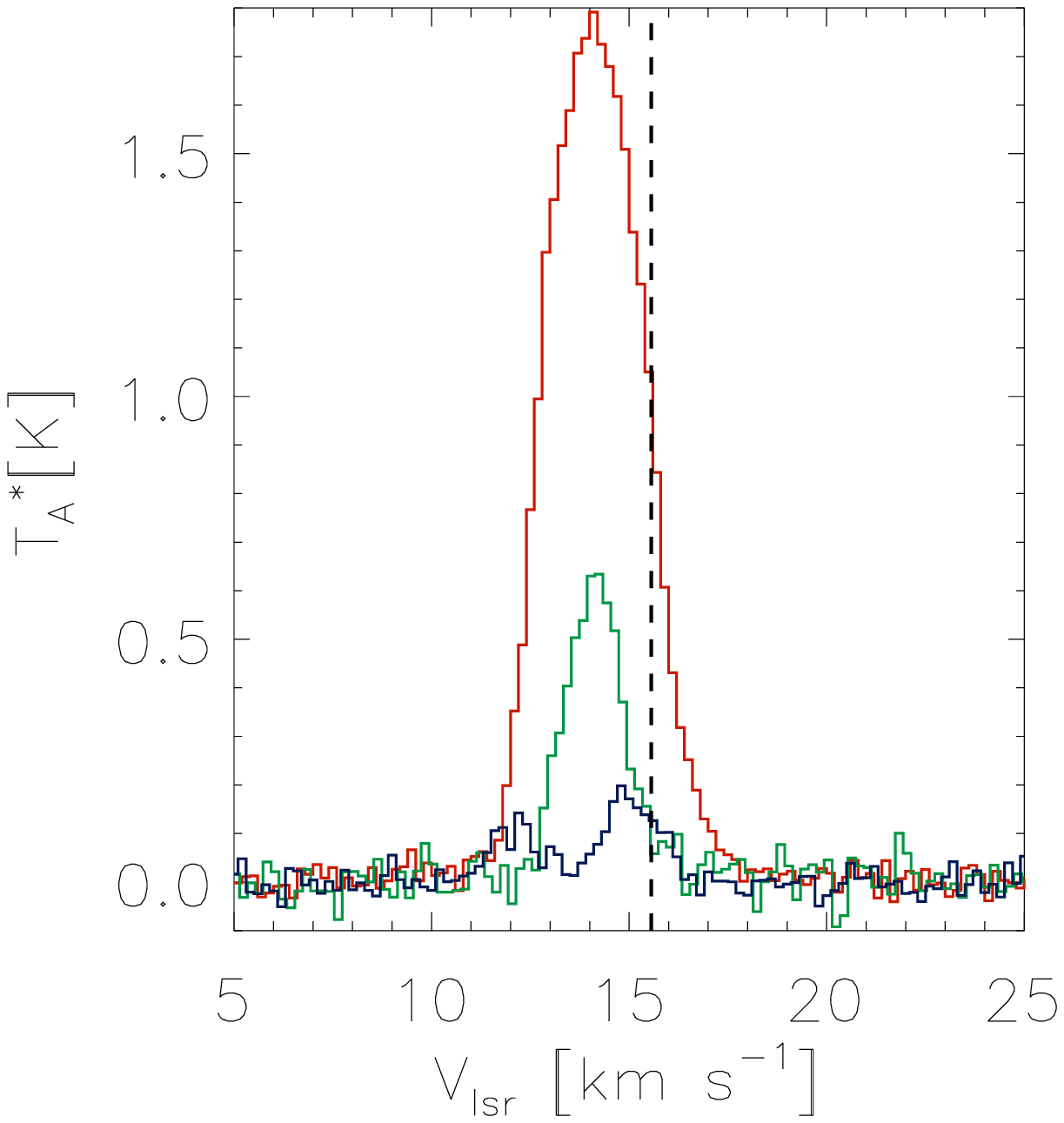}    
%
%
%
\caption{
Positionally averaged spectra of
$^{13}$CO (red), C$^{18}$O ( green) and C$^{17}$O$(1-0)$ (blue) lines.
The regions considered (from  left to right) are: a $\sim 4 \times 4\,$arcmin$^2$ region
around positions (RA $\sim 07^h11^m18^s$, DEC $\sim -10^{\circ}36'$)      
(corresponding to the isolated individual region in the SE);
a $\sim 4 \times 2\,$arcmin$^2$ region around position
(RA $\sim 07^h10^m42^s$, DEC $\sim -10^{\circ}23'$), at the NE end of the
secondary filament; and a $\sim 8 \times 3\,$arcmin$^2$ region around the
bright region of emission toward the NW along the main filament. These last two regions are
shown as black boxes in Fig.~\ref{fig:maps}. The dashed vertical line corresponds to the velocity, 15.6\,km\,s$^{-1}$
obtained with a Gaussian fit to the C$^{18}$O$(1-0)$ line of the first region.
}
\label{fig:avgspectraXS}
\end{figure*}

\subsection{Basic cloud structure: integrated intensity}
\label{sec:maps}

We present maps of the integrated intensity of the CO isotopologues in Fig.~\ref{fig:maps}.  
These maps were meant to cover most of the emission from the main and secondary filaments shown in 
Fig.~\ref{fig:map250}.
The main filament is visible as a strip of intense emission from the NW to SE. 
A secondary, less intense filament, can be observed running from the NE toward 
the center of the map where it meets the main filament. 
Although our coverage across the $\ell=224^{\circ}$ region is not complete, the observed
spectral line maps cover most of the warm dust emission from both main and secondary
filaments, as shown in Fig.~\ref{fig:map250}.  
As mentioned in Section~\ref{sec:mopra} we also mapped a smaller region,
visible in the SE corner of the maps and relatively off the main filament,
which contains a small group of starless clumps. 
However, this isolated tile will not be shown on subsequent maps, for
the sake of clarity.

Both $^{13}$CO and C$^{18}$O have a good signal-to-noise ratio (SNR) along the main
filament. However, the secondary filament is better traced by $^{12}$CO and $^{13}$CO.
One can see that in general the  $^{13}$CO and C$^{18}$O emission closely follows the
dust continuum emission, whereas the optically thick $^{12}$CO shows significant deviations, 
as expected.
The  most intense region of extended emission, located along the main filament  in the NW 
of the Hi-GAL map
(at approximately RA$\sim 07^h09^m19^s$ and DEC$\sim -10^{\circ}27'$), is clearly visible 
in the less optically thick $^{13}$CO and C$^{18}$O tracers.
%
However, we note a lack of correspondence between the C$^{18}$O emission and the dust continuum peaks 
which are visible near the center of the Hi-GAL map (at RA$\sim 07^h10^m05.8^s$ and DEC$\sim -10^{\circ}31'30''$).
C$^{18}$O and submillimeter dust continuum emission are both tracers of column density, 
but they have different dependences on temperature. Therefore, temperature variations, more likely to be observed 
towards dense clumps, may affect the relative distributions of these two column density tracers (see, e.g., \citealp{wilson1986}).
In addition, this departure between  the C$^{18}$O  and dust continuum  emission coincides with the location of 
several dense clumps and thus  another alternative (or additional) explanation for the lack of C$^{18}$O emission is that 
that this molecule has depleted from the dense gas (see, e.g., \citealp{hacar2013} and references therein).

%
\begin{figure*}
\centering
%
\includegraphics[width=18.0cm,angle=0]{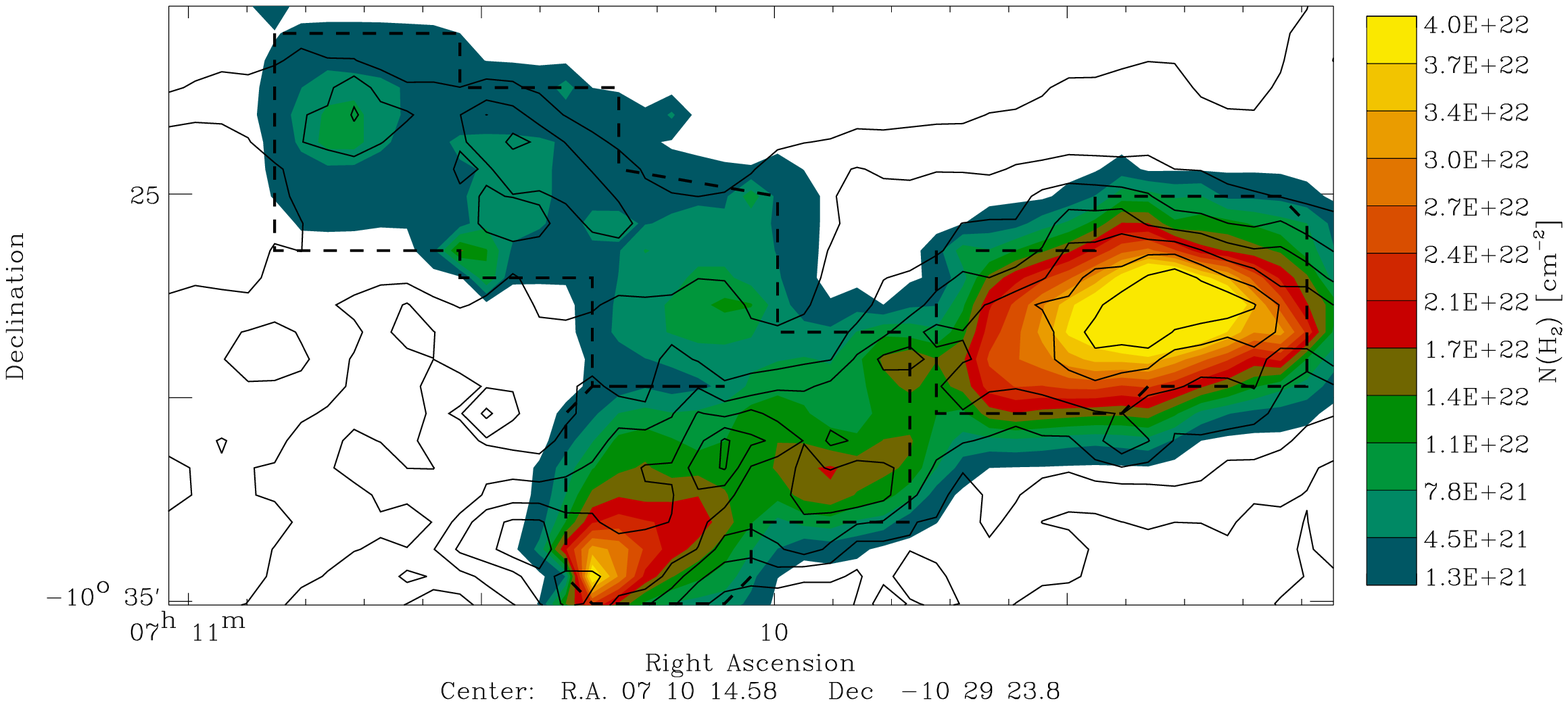}
\includegraphics[width=18.0cm,angle=0]{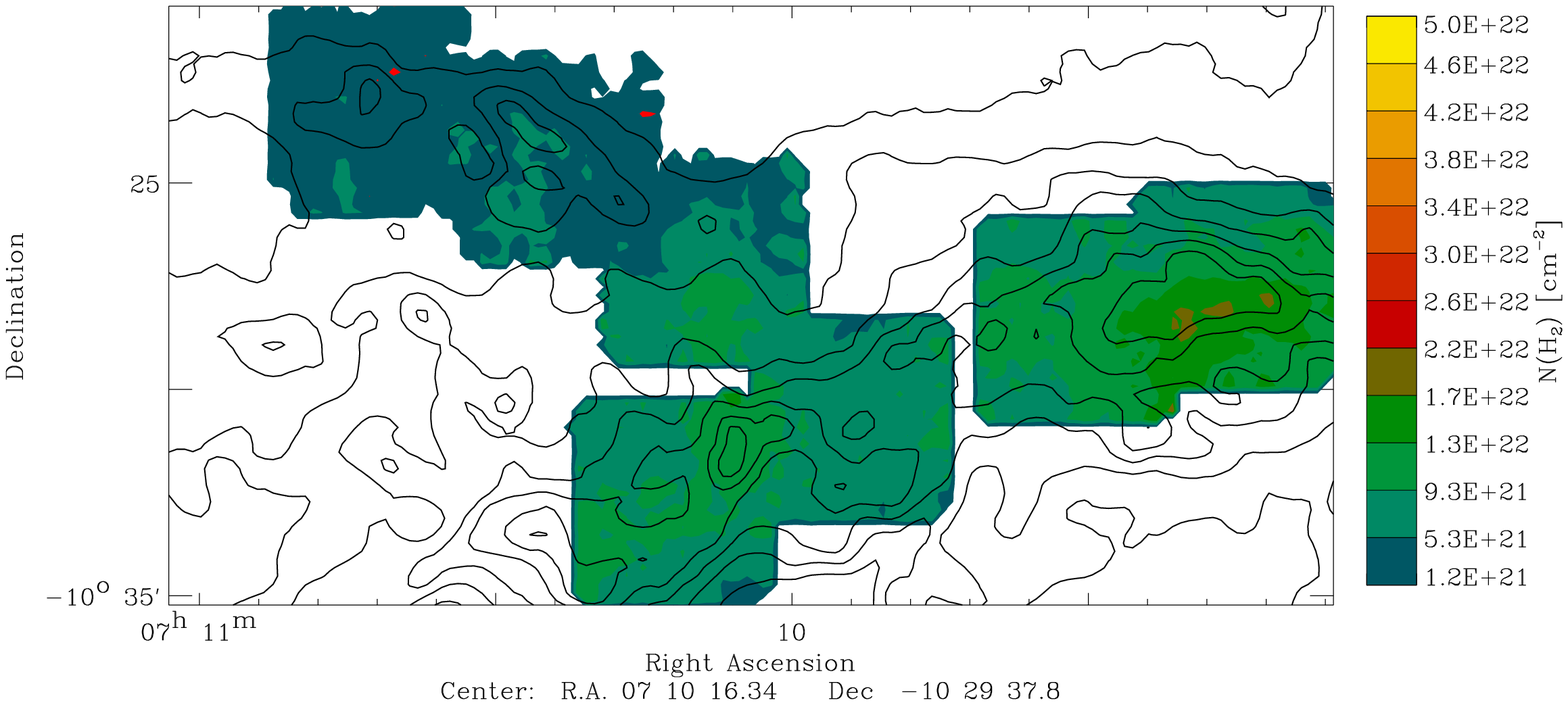}
\caption{
H$_2$ column density map as derived from the $^{13}$CO and C$^{18}$O$(1-0)$ data, regridded with 
the program {\tt xs} to a 80\,arcsec grid size and 40\,arcsec spacing (top panel),
and from the $^{13}$CO and $^{12}$CO$(1-0)$ data (bottom panel).  
Overlaid are the contours of the column density as derived by \citet{elia2013} 
(with first contour corresponding to  $2.8 \, 10^{21}\,$cm$^{-2}$  and subsequent contours 
equally spaced on a logarithmic scale, with Log$(\Delta N [{\rm cm}^{-2}]) = 0.2$), 
which are also regridded in the top panel.
Abundance ratios of [$^{13}$CO]/[C$^{18}$O]$\simeq 7$ and [C$^{18}$O] / [H$_2$] $\simeq 10^{-7}$
were assumed (see text).
}
\label{fig:cdC18O}
\end{figure*}

\subsection{Basic velocity structure: channel maps}
\label{sec:chmaps}

The area that we mapped with the Mopra telescope has already been covered by the 
lower angular resolution (2.7\,arcmin) survey of
molecular clouds in the Monoceros and Canis Major regions
carried out with the NANTEN telescope in the $^{13}$CO\,$(1-0)$ line
\citep{Kim2004}, spanning an area of 560\,deg$^2$. 
The NANTEN observations were also much less sensitive, with a typical RMS noise of $\sim 0.5$\,K.
However, although the NANTEN data can be used to estimate the kinematic distances needed 
to derive masses and luminosities \citep{elia2013}, higher angular resolution and more 
sensitive maps are necessary to analyze the column density distribution in the 
$\ell=224^{\circ}$ region.

We show the channel maps of the $^{13}$CO\ and C$^{18}$O lines in
Figures~\ref{fig:chmap13CO} and \ref{fig:chmapC18O}.
The distribution of bulk emission in the NW part of the cloud between $\sim 13.5$ and
16\,km\,s$^{-1}$ is similar in both tracers. As mentioned above, the emission toward
the center of the field is better detected in the $^{13}$CO map.
Extended emission from $^{13}$CO in both the main and secondary filaments is better viewed at
$V_{\rm lsr} \sim 14.2 - 14.6\,$km\,s$^{-1}$. Emission from  the secondary filament appears
to be concentrated on a narrower velocity range, compared to emission from the main
filament. Along the latter a velocity gradient (see Section~\ref{sec:velgrad}) is also visible, 
with emission slowly shifting from the NW (at low velocities) to the SE 
(at higher velocities). A similar velocity gradient is also observed in the 
C$^{18}$O channel map.

The averaged spectra of $^{13}$CO,  C$^{18}$O and C$^{17}$O 
are shown in Fig.~\ref{fig:avgspectraXS}.
The minor deviations from a Gaussian shape and the relatively large velocity range of emission
($\ga 5$\,km\,s$^{-1}$ for $^{13}$CO) are a consequence of the velocity gradients in the
region.  The $^{13}$CO emission from two of the regions considered is clearly double-peaked
(see also Figures~\ref{fig:chmap13CO} and  \ref{fig:chmap13COconv}).
This appears to be the consequence of two separate cloud components
at different velocities (14.4 and 16.2\,km\,s$^{-1}$),
rather than a dip due to self-absorption. In fact, the
average C$^{18}$O$(1-0)$ spectra from the same regions show weak emission to be
present at the lower-velocity component, which also appears to be the most intense.
However, the limited extent of our maps does not allow us to investigate whether this
$^{13}$CO double-peaked profile actually corresponds to two entirely different filaments 
that overlap in projection, as for example observed by \citet{hacar2013}. 
%

An interesting feature is the double-peaked profile of the C$^{17}$O spectrum toward
the main clump of emission. The more intense, high-velocity component corresponds to the same 
emission as from $^{13}$CO and C$^{18}$O. However, the lower-velocity component, peaking at 
about 12\,km\,s$^{-1}$,  has no correspondence in the spectra of the more abundant isotopologues,
and a clump of emission at this velocity can be clearly seen in Fig.~\ref{fig:chmapC17Oconv}.
Given the difference in the typical optical depth of C$^{17}$O and the other 
isotopologues, one possibility to explain these data is that the lower-velocity 
clump is behind the presumably more massive, high-velocity component, 
thus preventing detection of the less massive clump in the optically thicker 
$^{13}$CO and C$^{18}$O lines. 
Fig.~\ref{fig:chmapC17Oconv} also shows another interesting feature, namely, 
C$^{17}$O emission along the main filament is clearly detected at different velocities, peaking
for example in the 13.6\,km\,s$^{-1}$ channel and then rising again near 16\,km\,s$^{-1}$. 
This suggests that these filaments may indeed be characterized by a more complex 
velocity structure, as observed by \citet{hacar2013}.

\section{Analysis}
\label{sec:analysis}

%

In this section we will discuss the derivation of various physical parameters
such as column density, mass and kinetic temperature, that
are fundamental for the analysis of the sources. We will also analyse the kinematics
of the sources, determining velocity gradients and line asymmetries. The estimate of
the virial masses will also allow us to determine which sources are currently
gravitationally bound and which are not.

\subsection{Opacity, excitation temperature and column density}
\label{sec:coldens}

\subsubsection{Results from the $^{13}$CO and C$^{18}$O data}
\label{sec:C18O}

%

%
%
\begin{table*}[!ht]
\centering
\caption{Results of velocity gradient fitting.  }
\begin{tabular} {lccccccccccc}
\hline\hline
\noalign{\smallskip}
Line     & \multicolumn{3}{c}{\bf Whole region\tablefootmark{a}}   &   & \multicolumn{3}{c}{\bf Main filament}   &   & \multicolumn{3}{c}{\bf Secondary filament}  \\
\cline{2-4}
\cline{6-8}
\cline{10-12}
%
         & $V_o$            & ${\rm d}V/{\rm d}r$         & $\theta_v$\tablefootmark{b}    &
         & $V_o$            & ${\rm d}V/{\rm d}r$         & $\theta_v$\tablefootmark{b}    &
         & $V_o$            & ${\rm d}V/{\rm d}r$         & $\theta_v$\tablefootmark{b}    \\
\noalign{\smallskip}
         & [km\,s$^{-1}$]   & [km\,s$^{-1}$\,pc$^{-1}$]   &  [deg]   &   & [km\,s$^{-1}$]   & [km\,s$^{-1}$\,pc$^{-1}$]   &  [deg]
&   & [km\,s$^{-1}$]   & [km\,s$^{-1}$\,pc$^{-1}$]   &  [deg]  \\
\hline
\noalign{\smallskip}
%
C$^{18}$O\,$(1-0)$   & 15.1    & 0.42    & 292   &   & 15.3    & 0.35    & 318   &   & $-$       & $-$       & $-$ \\   
$^{13}$CO\,$(1-0)$   & 14.9    & 0.25    & 293   &   & 15.2    & 0.30    & 311   &   & 15.2      & 0.35      & 279 \\
\noalign{\smallskip}
\hline

\label{Table:velgrad}
\end{tabular}

\tablefoot{
\tablefoottext{a}{Including the isolated region in the SE.}
\tablefoottext{b}{The angle $\theta_v$ is measured positive from the axis of positive RA toward the north.}
}

\end{table*}
%

Column densities were derived from the $^{13}$CO and C$^{18}$O data by assuming a
filled beam and a uniform excitation temperature (equal for both tracers) within the beam.
We follow the standard procedure of deriving the opacity in both lines at each position
in the maps, from the ratio, $R$, of $^{13}$CO to C$^{18}$O  main-beam brightness temperatures:
\beq
R = \frac{T_{\rm mb}[^{13}{\rm CO}]} { T_{\rm mb}[{\rm C}^{18}{\rm O}] } =
\frac{ 1 - \exp(-\tau_{13}) } { 1 - \exp(-\tau_{18})   } \, .
\label{eq:tau}
\eeq
%
To reduce the effects of noise, the spectra were binned to a velocity resolution of
0.3\,km\,s$^{-1}$, and the brightness ratio was only computed when the SNR was $> 3$
in the C$^{18}$O  data. If we assume an abundance ratio [$^{13}$CO]/[C$^{18}$O]$\simeq 7.4$
\citep{wilson1994} then $\tau_{13} \simeq 7 \tau_{18}$ and Eq.~(\ref{eq:tau}) can thus be
solved for $\tau_{18}$. We calculate the opacity of the two CO isotopologues at the
peak of the C$^{18}$O spectrum at each position in the map.

If the optical depth is known, the excitation temperature $T_{\rm ex}$ can then be
derived easily from the equation of radiative transfer:
\beq
T_{\rm mb} = [ J_\nu(T_{\rm ex}) - J_\nu(T_{\rm bg}) ] \, [ 1 - \exp(-\tau) ] \, ,
\label{eq:radtransf}
\eeq
where we assumed a unity filling factor  and $T_{\rm bg} = 2.7\,$K. In Eq.~(\ref{eq:radtransf})
we also used $J_\nu(T) \equiv T_o / [ \exp(T_o/T) -1 ]$, with $T_o = h \nu / k$, where
$h$ and $k$ represent the Planck and Boltzmann constants, respectively.
%

%
\begin{figure*}
\centering
%
\includegraphics[width=18.0cm,angle=0]{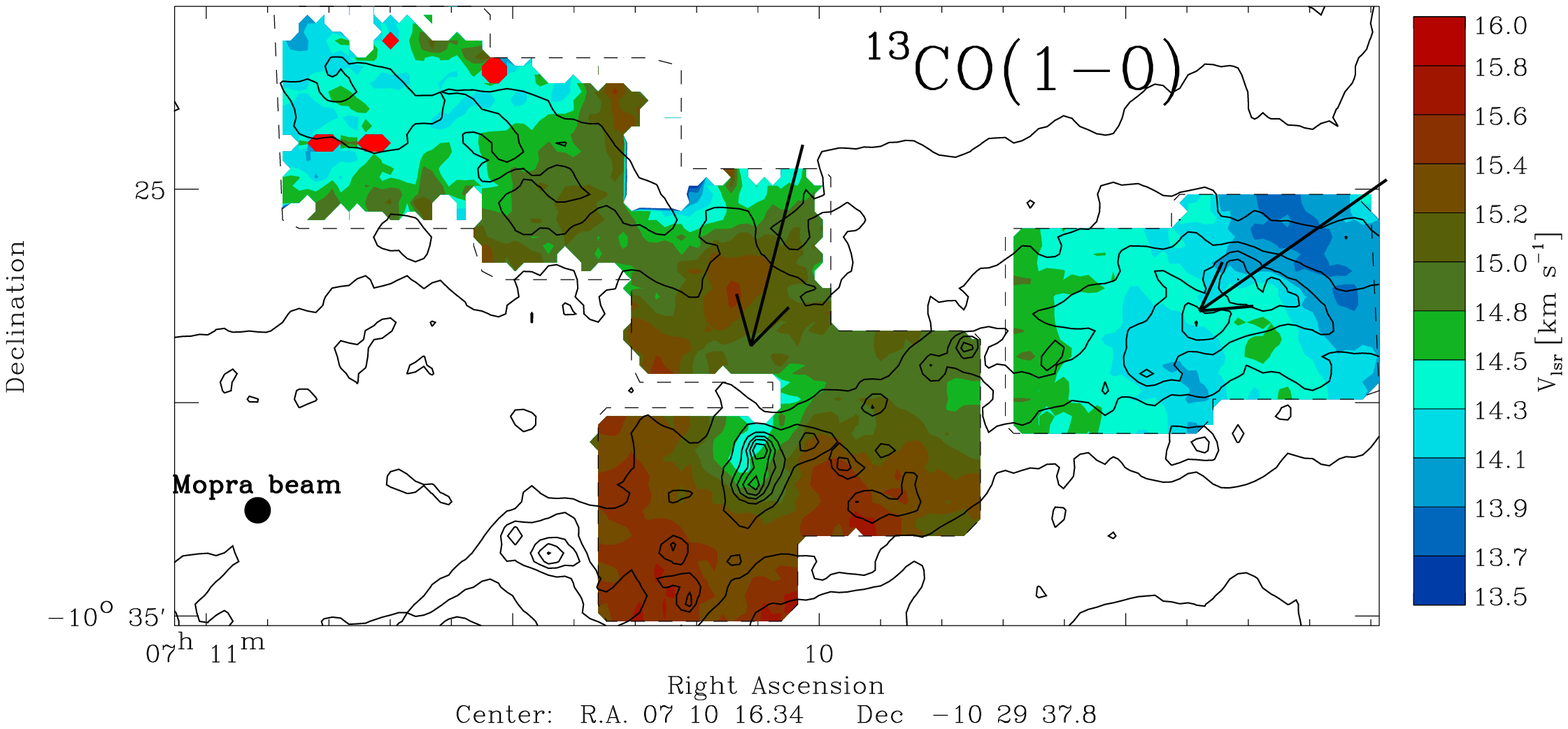}
\includegraphics[width=18.0cm,angle=0]{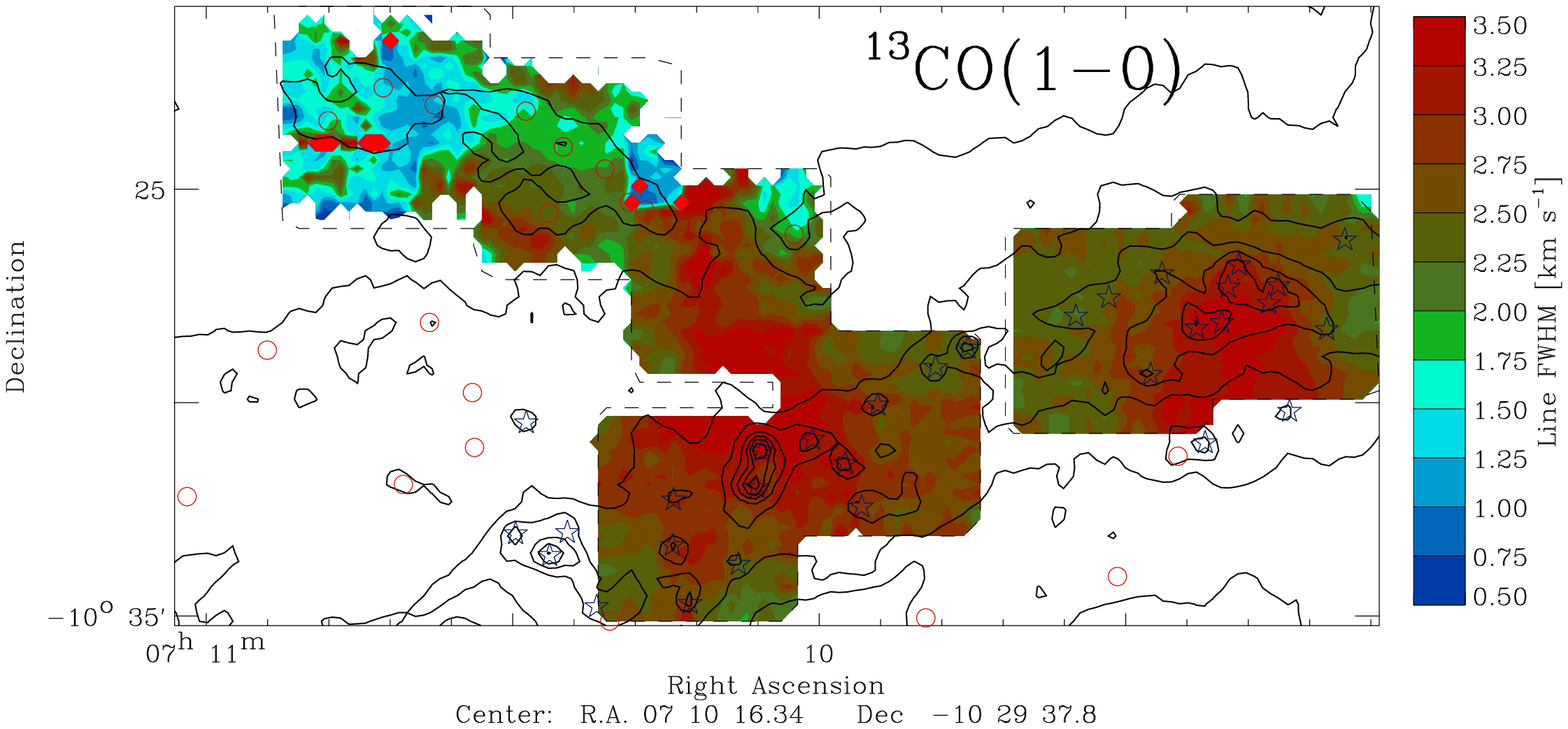}
\caption{
Kinematics toward the $\ell=224^{\circ}$ region. The top panel shows the line center
velocity from the $^{13}$CO$(1-0)$ data, while the bottom panel shows the line width.
The black arrows represent the direction and relative magnitude
of the velocity gradients along the main and secondary filaments (see Section~\ref{sec:velgrad}), 
while the black filled circle at the bottom left (top panel) indicates the Mopra beam FWHM.
Symbols in the lower panel are as in Fig.~\ref{fig:map250}.
The overlaid contours of the SPIRE $250\,\mu$m emission are as in Fig.~\ref{fig:cdC18O}.
The line centers and line widths have been computed by fitting a 1- or 2-component Gaussian
to the spectrum at each point (see text).  
}
\label{fig:kinematics13CO}
\end{figure*}

%
\begin{figure*}[!ht]      
\centering
%
\includegraphics[width=18.0cm,angle=0]{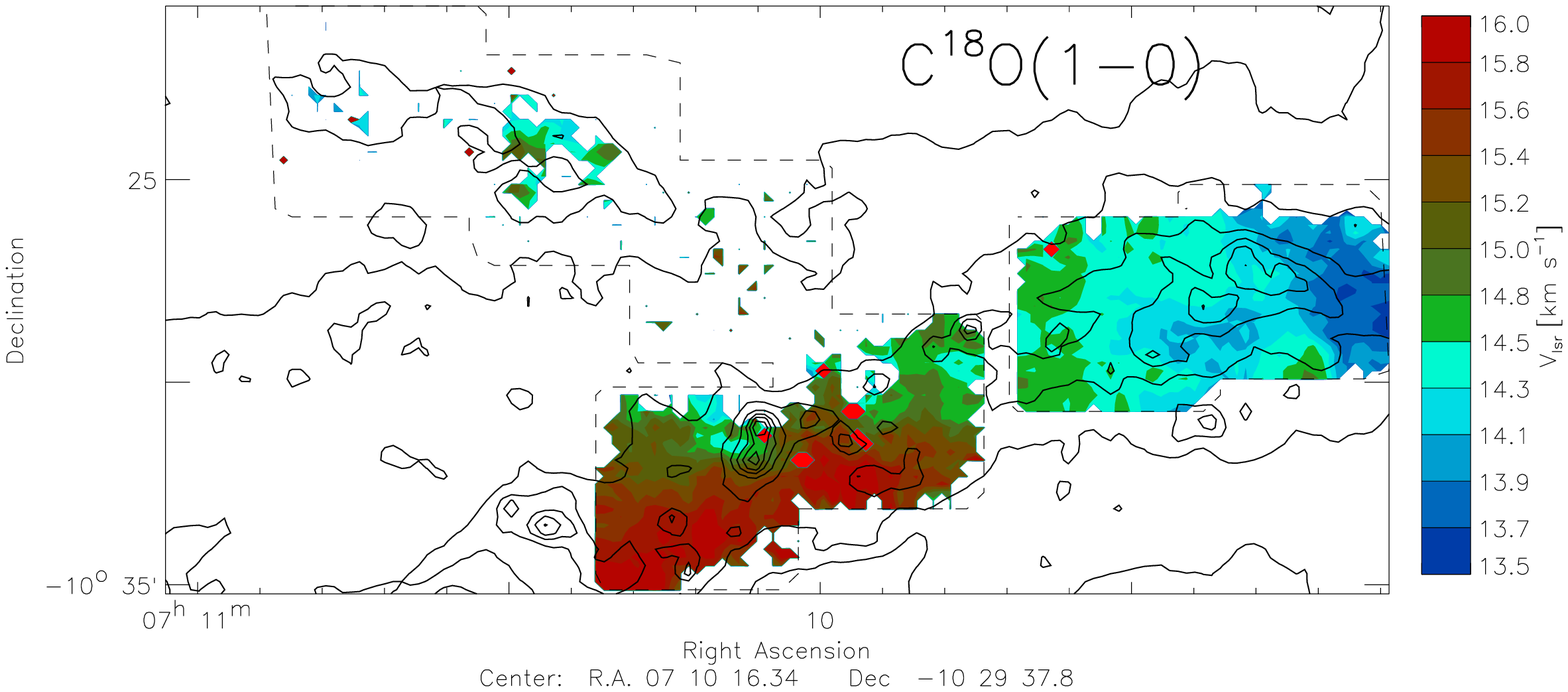}
\includegraphics[width=18.0cm,angle=0]{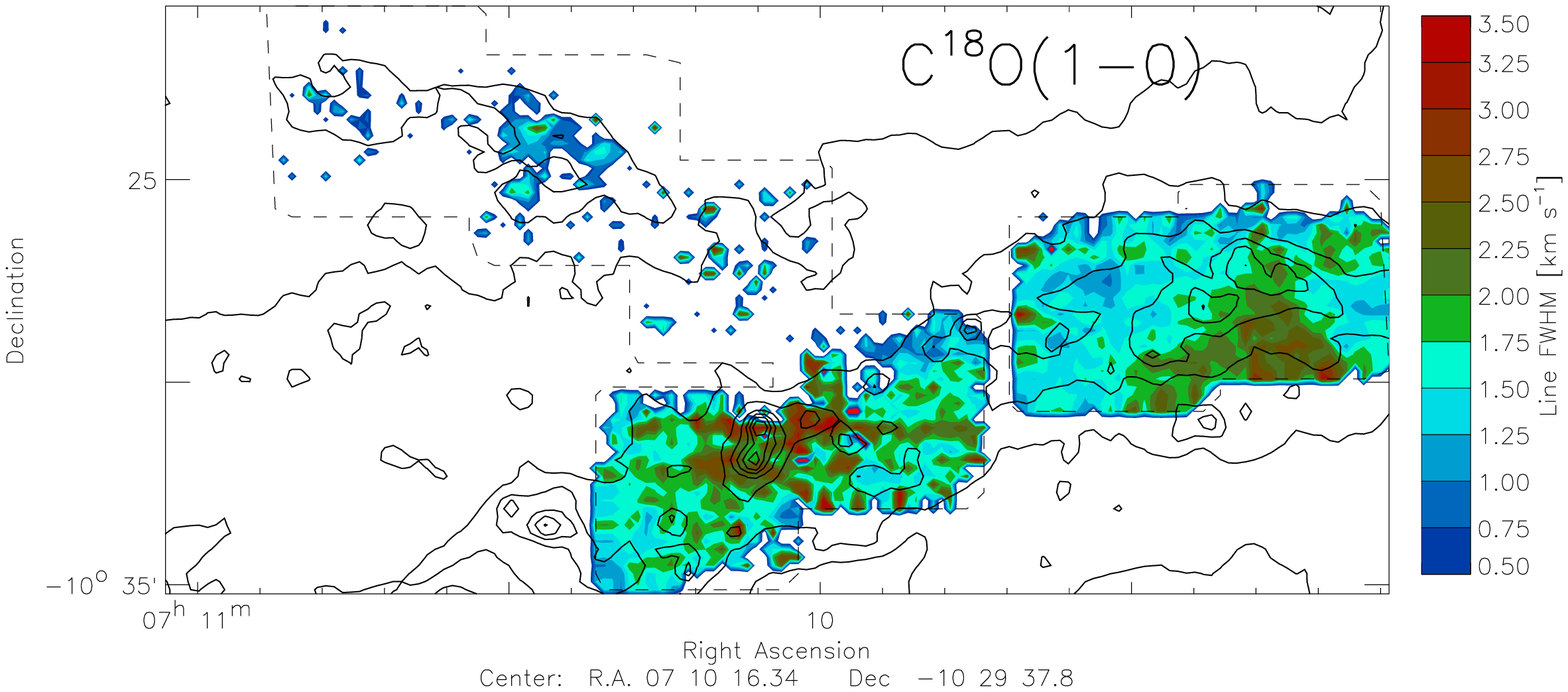}
\caption{
Same as Fig.~\ref{fig:kinematics13CO} for C$^{18}$O$(1-0)$.
The line centers and line widths have been computed by fitting a 1-component Gaussian
to the spectrum at each point (see text).
}
\label{fig:kinematicsC18O}
\end{figure*}

%
\begin{figure*}[!ht]
\centering
\includegraphics[width=9.0cm,angle=0]{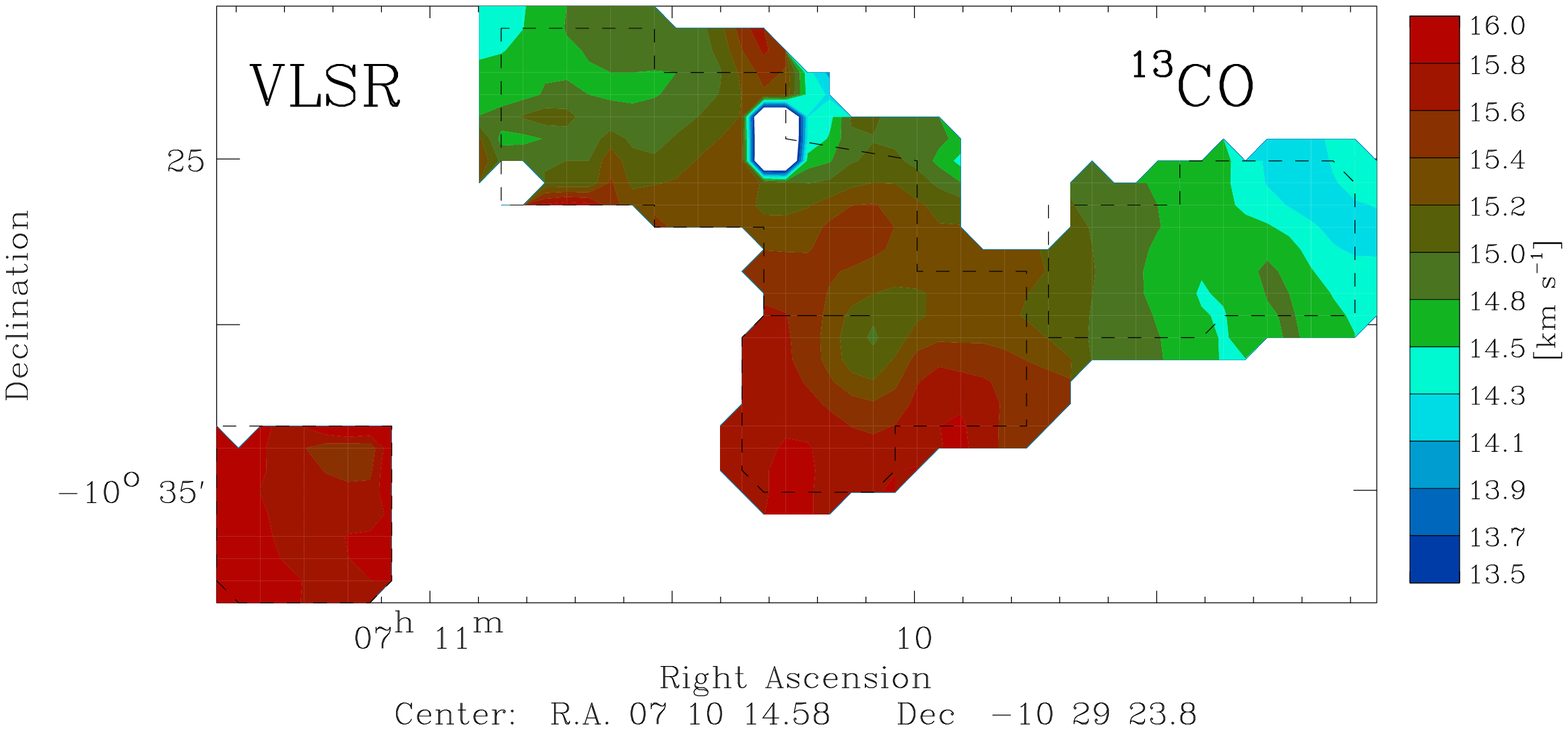}
\includegraphics[width=9.0cm,angle=0]{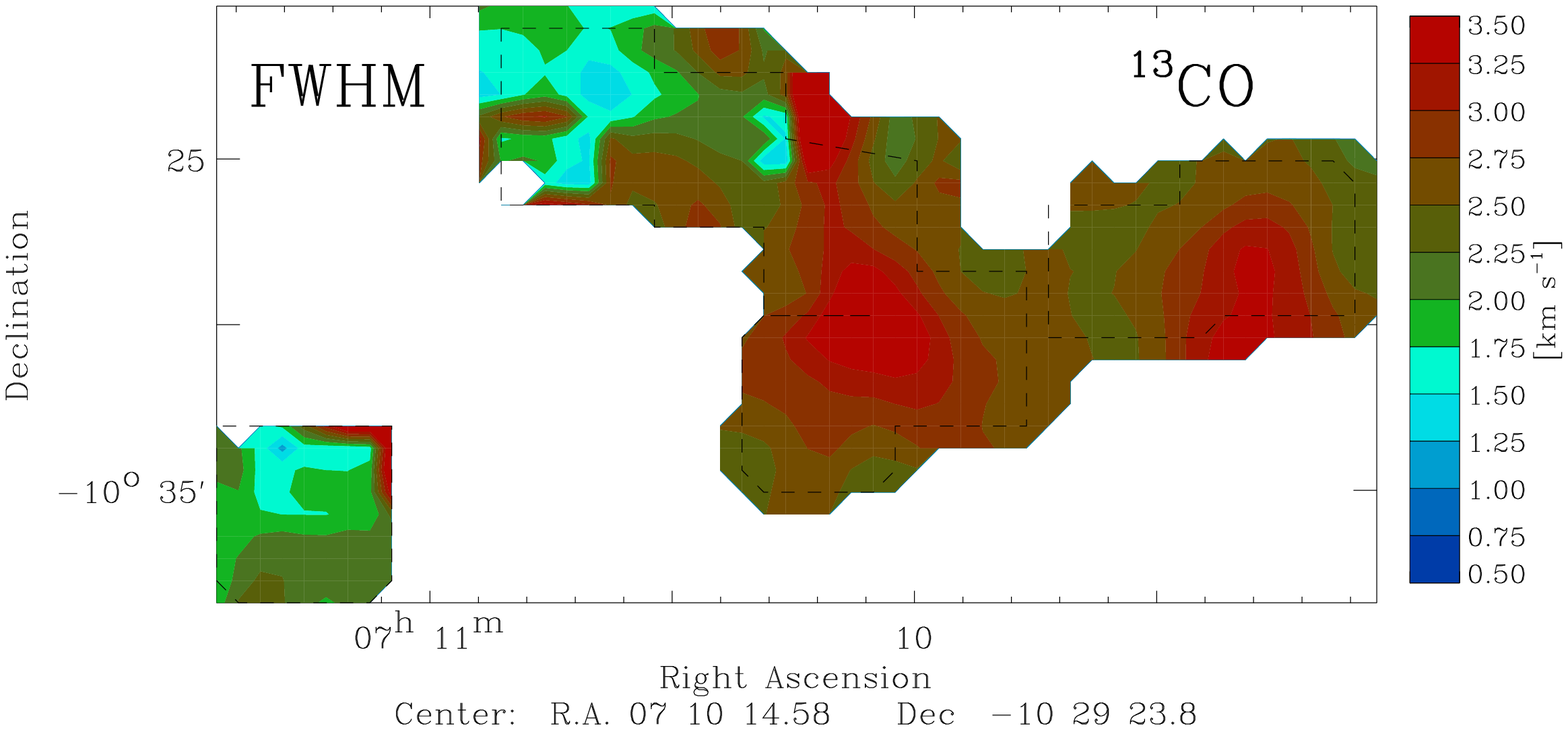} \\
\includegraphics[width=9.0cm,angle=0]{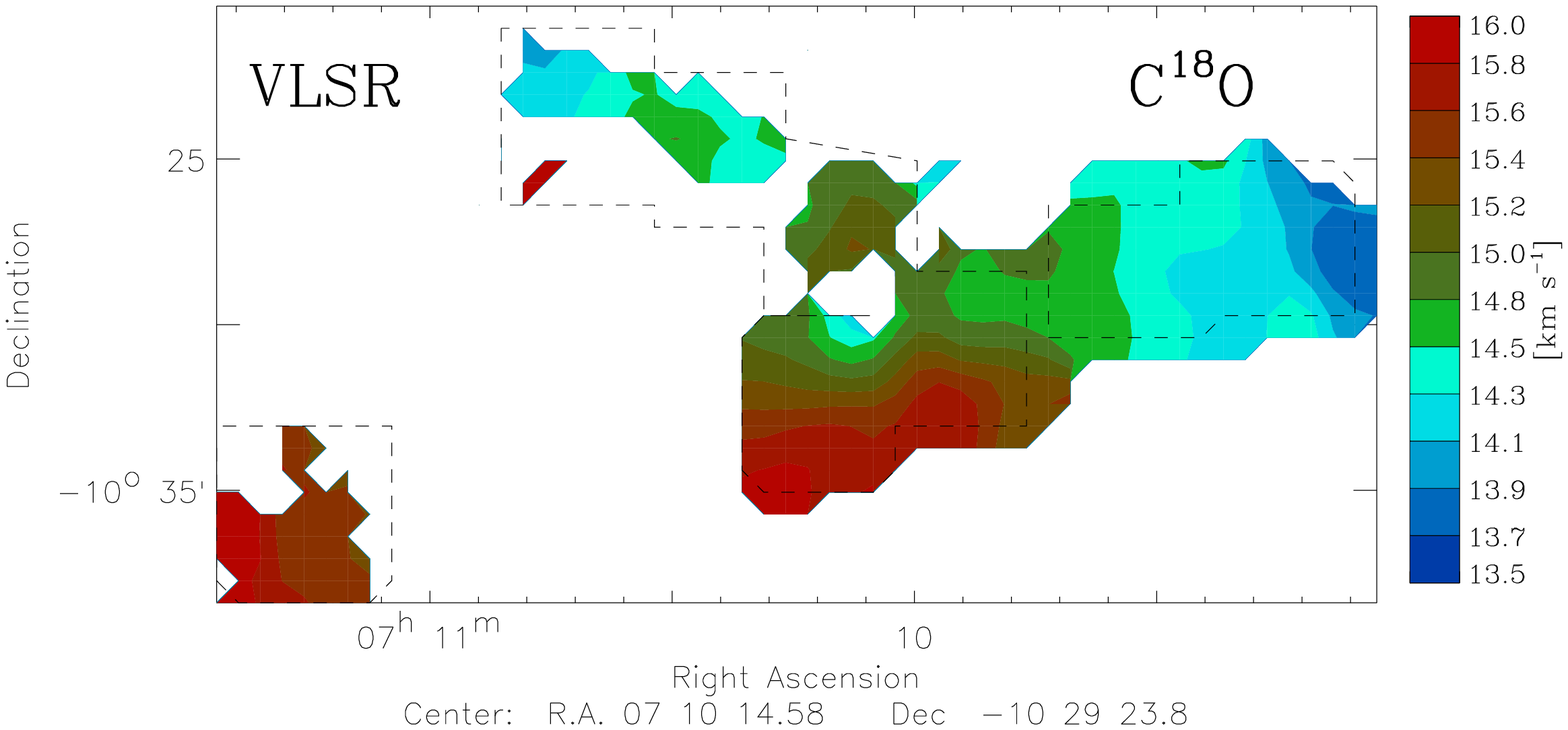}
\includegraphics[width=9.0cm,angle=0]{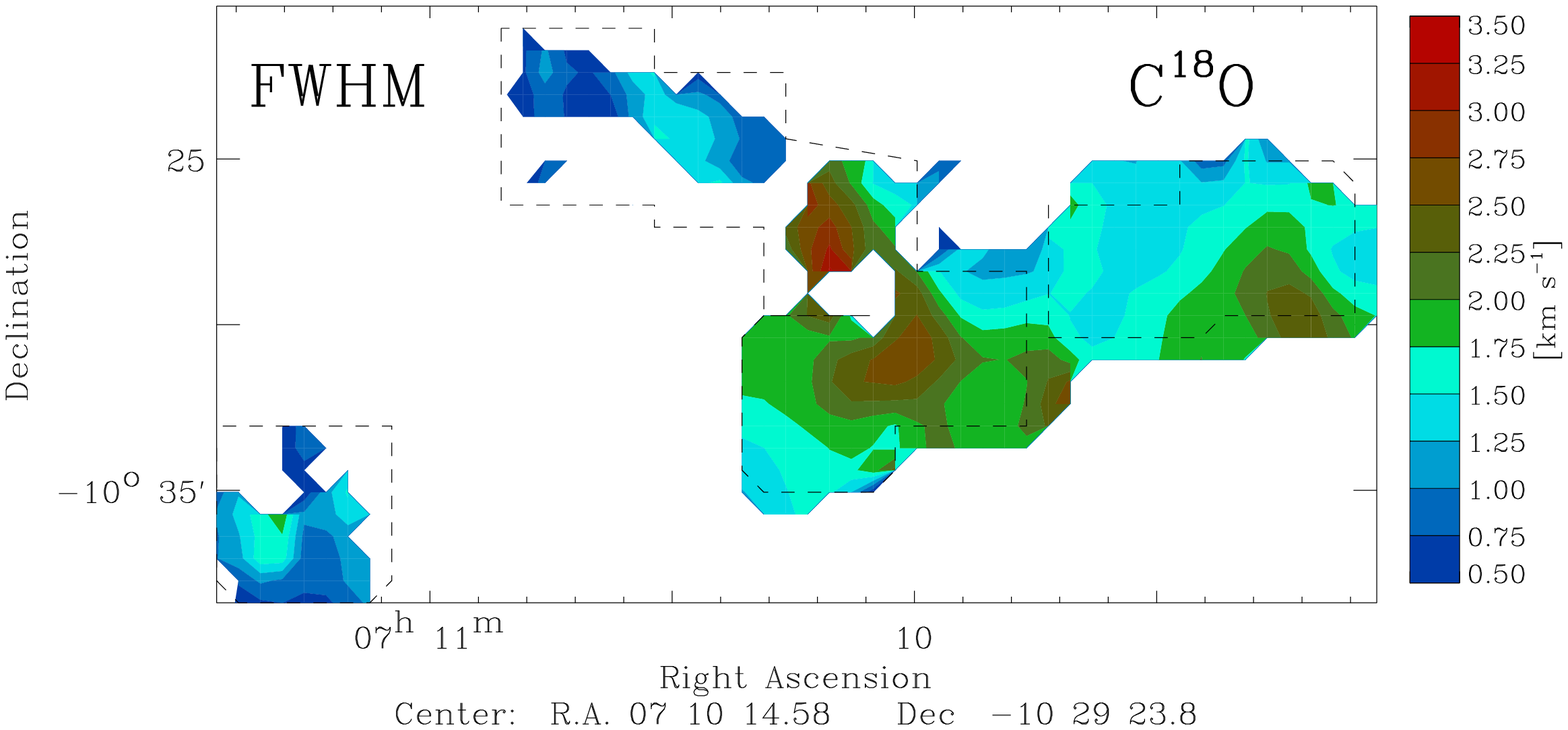} \\
\includegraphics[width=9.0cm,angle=0]{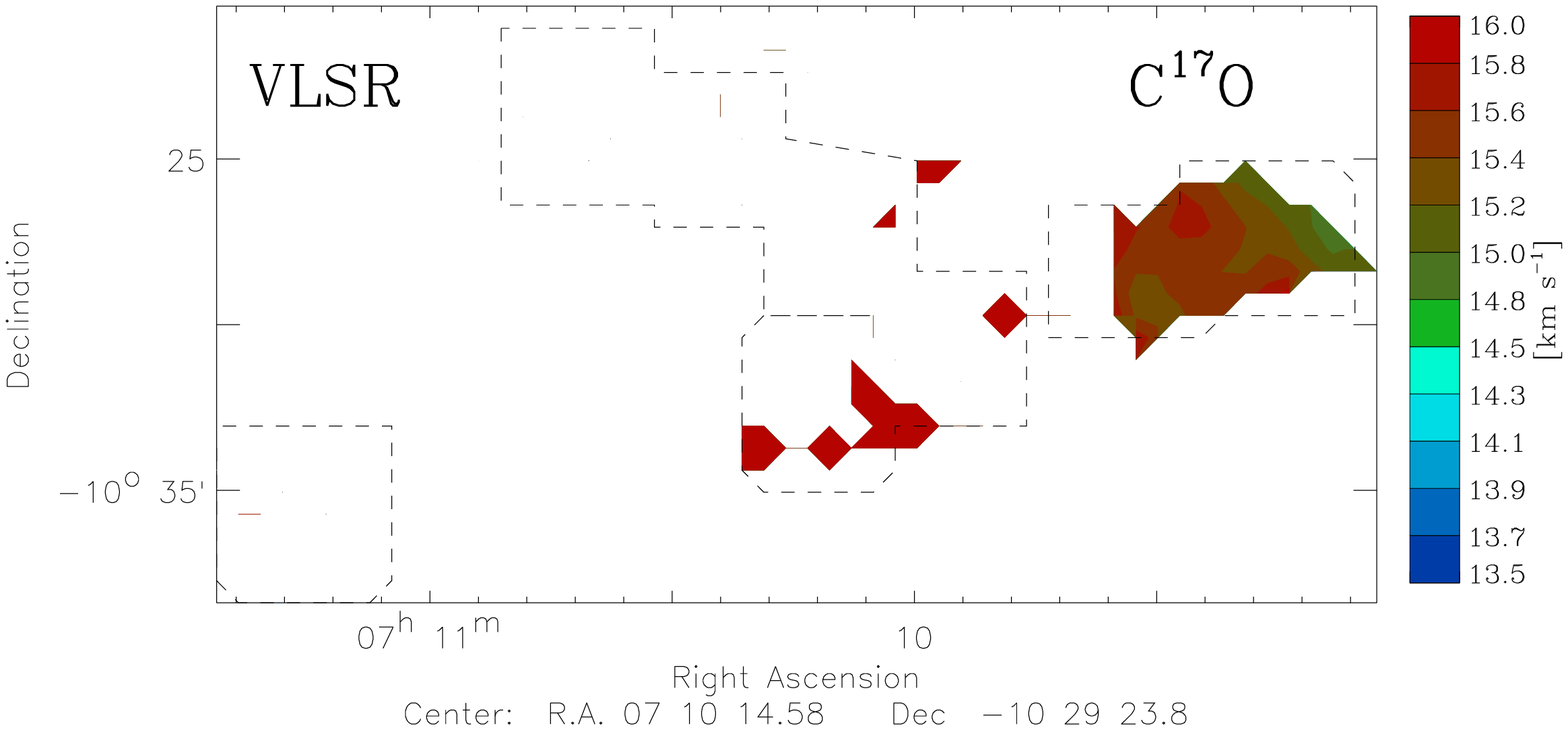}
\includegraphics[width=9.0cm,angle=0]{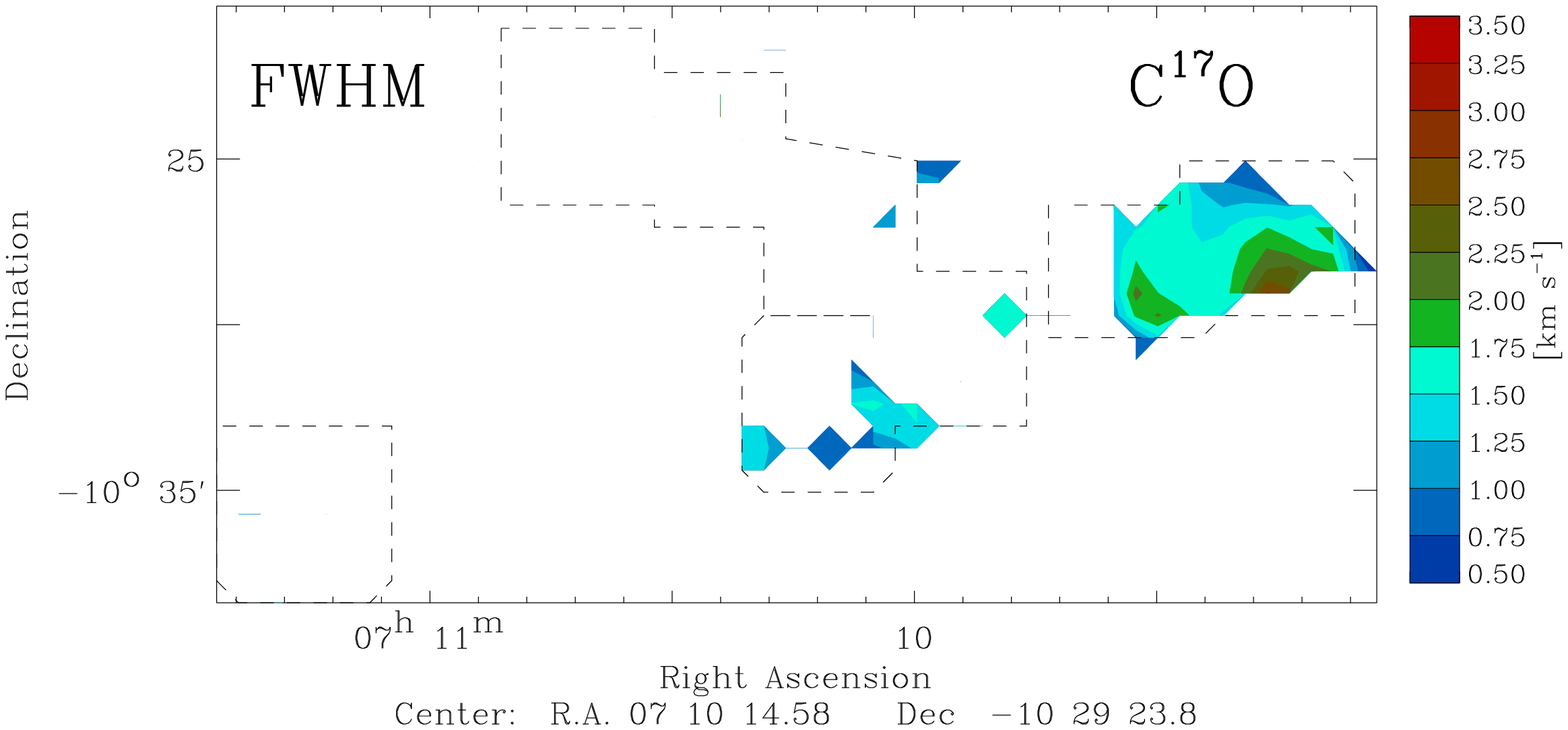}
\caption{
Line center velocities (left column) and line widths (right column)  after convolving (from top to bottom)
the $^{13}$CO, C$^{18}$O and C$^{17}$O$(1-0)$ Mopra maps
to 80\arcsec and regridding with a 40\arcsec spacing. The smaller red areas are map artifacts. 
}
\label{fig:kinematicsCONV}
\end{figure*}

Given $T_{\rm ex}$ and $\tau_{13}$ for the $J = 1 - 0$ transition,
the total column density, $N_{\rm mol}$, of $^{13}$CO can be derived using
the following formula (see, e.g., \citealp{lis1991}):
\begin{eqnarray}
& & N_{\rm mol} \, [{\rm cm}^{-2}] = \frac{4.0\times 10^{12}}
{J^2\mu^2[{\rm D}] \,B [{\rm K}] }
Z \exp\left( \frac{E_{\rm J}}{T_{\rm ex}} \right ) \times  \nonumber \\
& & \frac{1}{\eta_{\rm mb}} \frac{\tau}{1-e^{-\tau}}
\int \, T_{\rm A}^{\star} \, {\rm d}v \, [{\rm K~ km~ s}^{-1}] \, ,
\label{eq:cd}
\end{eqnarray}
where $B$ denotes the rotational constant, $E_{\rm J}$ is the
upper state energy, $\mu$ is the dipole moment  (in Debye) and we used the
escape probability $\tau/[1-\exp(-\tau)]$ to account for first-order
optical depth effects.                     
With $Z$ we have indicated the partition function of a linear molecule which, for
$k T_{\rm ex} >> h B$, is given by:
\begin{equation}
Z = \frac{k T_{\rm ex}}{h B} \, .
\end{equation}
The molecular hydrogen column density, $N({\rm H}_2)$, was then calculated assuming a
[$^{13}$CO] / [H$_2$] $\simeq 1.4 \times 10^{-6}$ abundance ratio \citep{frerking1982} 
and thus we obtain [C$^{18}$O] / [H$_2$] $\simeq 2 \times 10^{-7}$.
The top panel of Fig.~\ref{fig:cdC18O} 
shows the resulting column density map at all positions where both $T_{\rm ex}$ and $\tau_{13}$
could be reliably calculated. The column density and Hi-GAL $250\,\mu$m emission maps look quite
similar, with the notable exception of the enhanced $250\,\mu$m emission at position
(RA,DEC)$\simeq (07:10:05, -10:32:00)$ which has no corresponding clump in the $N({\rm H}_2)$ map.
This may be a consequence of C$^{18}$O not tracing closely the dust emission at this
position (see Sect.~\ref{sec:maps}) and also because of the partially missing data near this
position as well (see Figures~\ref{fig:maps} and \ref{fig:chmapC18O}). The missing data near
the peaks of $N({\rm H}_2)$ are due to the $^{13}$CO line becoming self-absorbed and
the ratio $R$ becoming $< 1$.
The typical values for excitation temperature and opacity across the measured region are
$T_{\rm ex} \sim 5 - 9$\,K and $\tau_{13} \sim 2 - 10$. This suggests that the $^{13}$CO line
is mostly optically thick, but also the C$^{18}$O line could be moderately optically thick
at some positions.

\subsubsection{Results from the $^{12}$CO and $^{13}$CO data}
\label{sec:12CO}

The $^{13}$CO and C$^{18}$O data are effective in deriving excitation temperature,
opacity and column density along the main filament. However, as shown by Fig.~\ref{fig:cdC18O},
there are regions in the map where the physical parameters of the gas could not be determined,
mainly because the C$^{18}$O spectra were not good enough. We then decided to use
$^{12}$CO and $^{13}$CO to extend the derivation of the physical parameters to most of the
field area mapped with the Mopra telescope.

Then, assuming that the $^{12}$CO transition is optically thick and that $T_{\rm pk}[^{12}$CO]
is the main beam brightness temperature at the peak of $^{12}$CO$(1-0)$, we can derive the excitation
temperature using Eq.~(\ref{eq:radtransf}):
\beq
T_{\rm ex} = \frac{ T_o[^{12}{\rm CO}] }
{ \ln[ 1 + T_o[^{12}{\rm CO}]/( T_{\rm pk}[^{12}{\rm CO}] + 0.86{\rm K} ) ] } \, ,
\label{eq:tex}
\eeq
where $T_o[^{12}$CO$] = h \nu[^{12}$CO$] / k = 5.5\,$K.

Then, assuming that the excitation temperatures of the $^{12}$CO and $^{13}$CO lines
are the same, the optical depth of $^{13}$CO can also be derived from Eq.~(\ref{eq:radtransf}):
\beq
\tau[^{13}{\rm CO}] = - \ln \left [ 1 - \frac{ T_{\rm pk}[^{13}{\rm CO}]/T_o[^{13}{\rm CO}]  }
{ 1/(\exp(T_o[^{13}{\rm CO}]/T_{\rm ex}) -1) - 0.16 } 
\right ] \, ,
\eeq
where $T_o[^{13}$CO$] = h \nu[^{13}$CO$] / k = 5.3\,$K and $T_{\rm pk}[^{13}{\rm CO}]$ is the
main beam brightness temperature at the peak of $^{13}$CO.
The main {\it caveat} of this method is that the most dense regions will not be accurately
probed by $^{12}$CO, whose emission is restricted to outer cloud material due to its higher
opacity.
The bottom panel of Fig.~\ref{fig:cdC18O} shows the resulting column density. The 
discrepancies with respect to the column density map obtained as described in Section~\ref{sec:C18O}
are clearly visible, and they are likely a consequence of the assumptions made for each method, 
in particular regarding the molecular abundances. We also note that the values for the H$_2$ column density
obtained by \citet{elia2013} using the dust continuum emission are typically intermediate between those 
shown in the lower and upper panels of Fig.~\ref{fig:cdC18O}, indicating an overall internal consistency. 

\subsection{Gas kinematics}
\label{sec:kinematics}

\subsubsection{Velocity gradients}
\label{sec:velgrad}

Systematic variations in the $^{13}$CO and C$^{18}$O line center velocity are apparent across
the mapped region.
In fact, it can be seen in Figures~\ref{fig:kinematics13CO}  to \ref{fig:kinematicsCONV}  that
the velocity gradually increases along a particular direction.

We measure the velocity gradient across the whole mapped region following the procedure outlined in
\citet{goodman1993}. We assume the centroid velocities of the spectral lines follow a simple linear form and
fit the function:
\beq
V_{\rm lsr} = V_o + a\, \Delta \alpha + b \, \Delta \delta \, ,
\eeq
to the peak velocity of a Gaussian fit to the emission profile of the
$^{13}$CO and C$^{18}$O lines.  Here, $\Delta \alpha$ and $\Delta \delta$
represent offsets in right ascension and declination, expressed in radians, $a$ and $b$
are the projections of the gradient per radian on the $\alpha$ and $\delta$
axes, respectively, and $V_o$ is the LSR systemic velocity of the cloud.

We found the best-fitting values to the constants $a$ and $b$ using MPFIT \citep{markwardt2009},
and then converted all of the angular velocity gradients to physical scale assuming a distance of 1\,kpc.
This yields the overall velocity gradients listed in Table~\ref{Table:velgrad}.
Since in the Gaussian fitting procedure only spectra with a SNR\,$\ge 4$ are considered, the velocity
gradient estimated from C$^{18}$O is mainly sensitive to systematic variations along the
main filament, while the velocity gradient estimated from $^{13}$CO is sensitive to velocity
variations across the whole mapped region. This may explain the higher value of ${\rm d}V/{\rm d}r$
derived from C$^{18}$O. In Table~\ref{Table:velgrad} the velocity gradient from C$^{18}$O along the
secondary filament has not been estimated because we have few spectra with SNR\,$\ge 4$.

The $V_o$ and $\theta_v$ parameters of the best-fit velocity gradients are 
quite similar for both molecular tracers. The values estimated from $^{13}$CO 
show that the directions of the velocity gradient along the main and secondary 
filament are clearly different, by more than 30\,deg. In particular,
the velocity gradient along the main filament is oriented almost parallel to the filament itself,
with velocity increasing roughly from the NE to the SE, as
also shown by Figures~\ref{fig:kinematics13CO} and \ref{fig:kinematicsC18O}.
The velocity gradient associated with the secondary filament appears instead almost
perpendicular to the filament itself. However, this result is less reliable because of
the limited extent of the mapped region, especially in the direction orthogonal to
the filament.

In Section~\ref{sec:evolution} we suggest that a possible physical cause for the observed
velocity gradients is the presence of accretion flows. However, at present we cannot exclude 
that the observed velocity gradients are produced by other motions.
For example, the velocity gradients could be interpreted as rotation, though 
our limited and irregular mapped area makes this interpretation more difficult. 
Then following \citet{goodman1993} we   
can calculate the parameter $\beta$, defined as the ratio of rotational kinetic energy to gravitational energy
for a sphere of uniform density $ \delta_\circ$:
%
%
\beq
\beta = \frac{\omega^2}  { 4 \pi G \delta_\circ} \, ,
\eeq
where $\omega$ is the angular velocity and $G$ is the gravitational constant. For the most intense region of emission
in the NW section of the main filament (shown as a black box in Fig.~\ref{fig:maps}) we find $\beta \simeq 0.08$, 
suggesting that if this interpretation were correct then
the rotational energy would be a negligible fraction of the gravitational energy.

At present, we do not have data indicating the presence of proto-stellar outflows in the mapped region. However, 
while outflows could be a plausible physical explanation for velocity gradients observed at the clumps scale,
it is unlikely that a single outflow or multiply oriented outflows could explain the observed large-scale velocity
gradients.

%
%
%
\begin{table}    
\centering
\caption{Line FWHM of the molecule of mean mass toward the Hi-GAL clumps (see Section~\ref{sec:virial}).  }
\begin{tabular} {lccccc}
\hline\hline
\noalign{\smallskip}
Clump type   & Clump center   & Ambient gas  \\
\noalign{\smallskip}
             & [km\,s$^{-1}$]    & [km\,s$^{-1}$]    \\
\noalign{\smallskip}
\hline
\noalign{\smallskip}
Starless          & $1.3 \pm 0.3$   & $1.4 \pm 0.5$  \\
Proto-stellar     & $1.8 \pm 0.5$   & $1.9 \pm 0.3$  \\
\noalign{\smallskip}
\hline

\label{Table:avglinewidth}
\end{tabular}

\end{table}
%

\subsubsection{Line profile and line widths}
\label{sec:linewidths}

Figures~\ref{fig:map250} and \ref{fig:cdC18O} show a reasonable agreement between the
distributions of dust emission and gas column density. Fig.~\ref{fig:map250} also 
shows that the Hi-GAL compact clumps concentrate in regions of higher column density
relative to the surrounding regions, as expected.
Figures~\ref{fig:kinematics13CO} and \ref{fig:kinematicsC18O} then show that 
on the main filament the column density roughly correlates also with the C$^{18}$O and 
$^{13}$CO velocity dispersion. However, Fig.~\ref{fig:kinematics13CO} shows that 
on the secondary filament such a correlation is much
less evident, and actually looks like an anti-correlation at some positions.

%
%
%

Likewise, Figures~\ref{fig:kinematics13CO} to \ref{fig:kinematicsCONV} show that starless 
clumps are preferentially located at positions where the line widths are generally 
lower than those observed at the positions of the protostellar clumps. 
This is confirmed by extracting the C$^{18}$O 
line width at the nominal position of each Hi-GAL clump, if the line was detected. 
Table~\ref{Table:avglinewidth} then shows the average line widths as measured from 
the C$^{18}$O spectral line cube at the original angular resolution (column ``Clump center''),
and also the average values after the cube has been spatially convolved (regridded with the program {\tt xs} to
a 80\,arcsec grid size and 40\,arcsec spacing). The latter values (column ``Ambient gas'') are more
representative of the molecular gas surrounding the clump, and should give a better estimate of the kinematical
status of the ambient gas immediately around the Hi-GAL clump. Table~\ref{Table:avglinewidth} shows that
in both case the C$^{18}$O line widths are larger toward the protostellar clumps, though they are
marginally consistent within the errors.
The location of starless clumps in regions of lower velocity dispersion is further confirmed by the small
linewidths measured in the isolated tile at the SE (see Fig.~\ref{fig:kinematicsCONV}), which hosts a small 
cluster of starless clumps.

Our analysis shows evidence of gas flow along the main filament. However, we do not find 
any signs of accretion flow from the filament onto the Hi-GAL clumps. In fact,
as mentioned in Section~\ref{sec:chmaps} the line profiles of our emission lines are
mostly Gaussian and double-peaked in some cases, and we do not observe clear signs
of self-absorption. The lack of (detectable) self-absorption means that we do not see
any sign of infall or gas accretion across the filaments. 
Self-absorbed line profiles in filamentary regions were instead observed, for example, 
toward the Serpens South cluster and filament \citep{kirk2013} and also toward the
high-mass star-forming region DR21 \citep{schneider2010}, where evidence for material
radially contracting onto the filaments has been found. In these two regions the
authors used the HCO$^+ \, (1-0)$ line to detect and analyze the self-absorption.
However, \citet{arzoumanian2013} did not detect any sign of self-absorption 
in their observations of C$^{18}$O\,$(1-0)$, $(2-1)$ and $^{13}$CO\,$(2-1)$ toward 
several filaments.
Clearly, the exact line profiles of HCO$^+$, C$^{18}$O and $^{13}$CO depend 
on the velocity and density structure of both the ambient gas and the molecular clumps.
Carrying out a detailed comparison of the line profiles of these different molecular tracers 
is beyond the scope of this paper.  However, even without using complex radiative transfer codes, 
it is possible to analyze the effects of various source and telescope parameters on the hypothetical 
blue-asymmetric infall line profile using an analytical model, such as the one discussed by \citet{devries2005}.
%

%
\begin{figure}[!ht]
\centering
\includegraphics[width=9.5cm,angle=0]{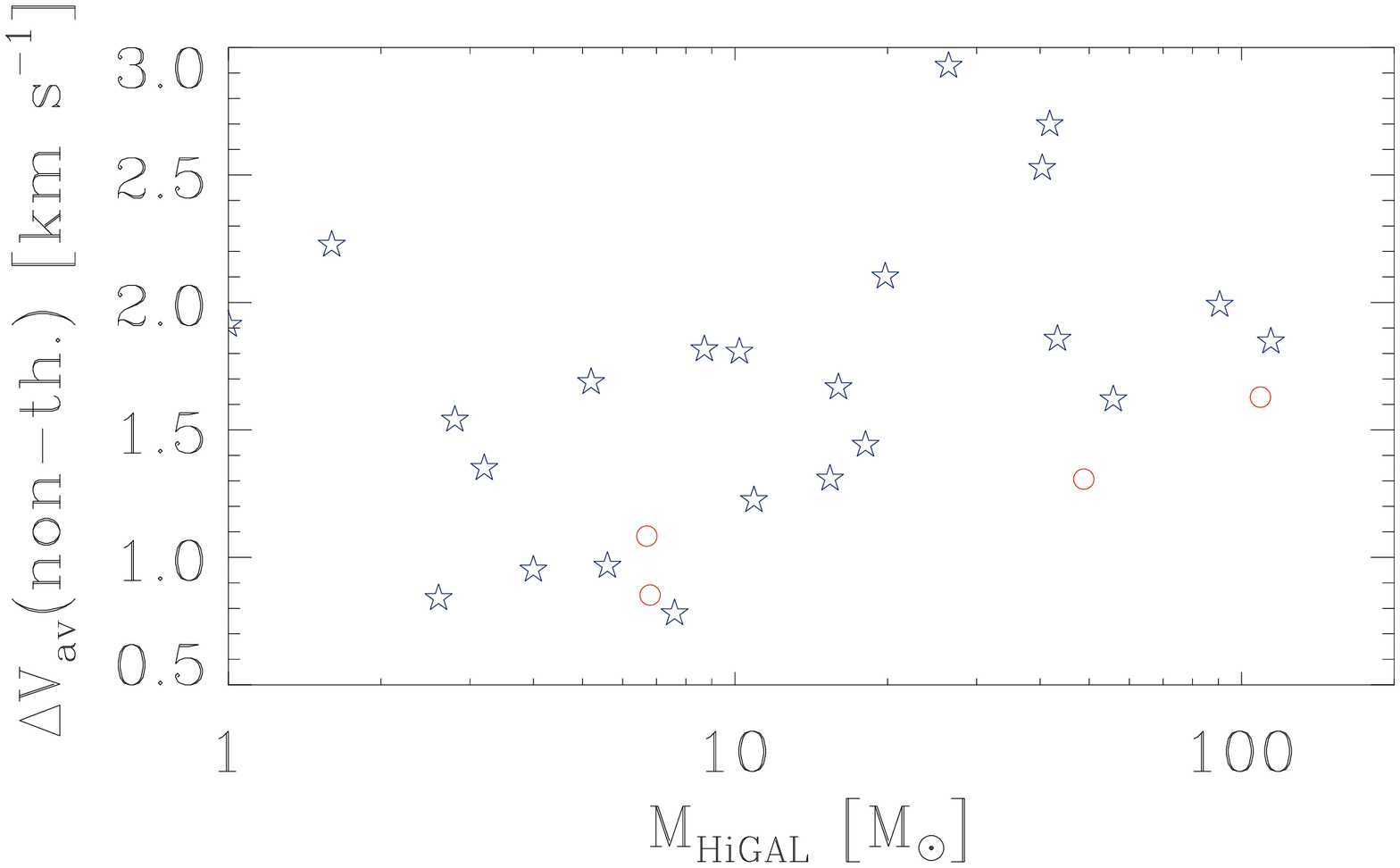}
\includegraphics[width=9.5cm,angle=0]{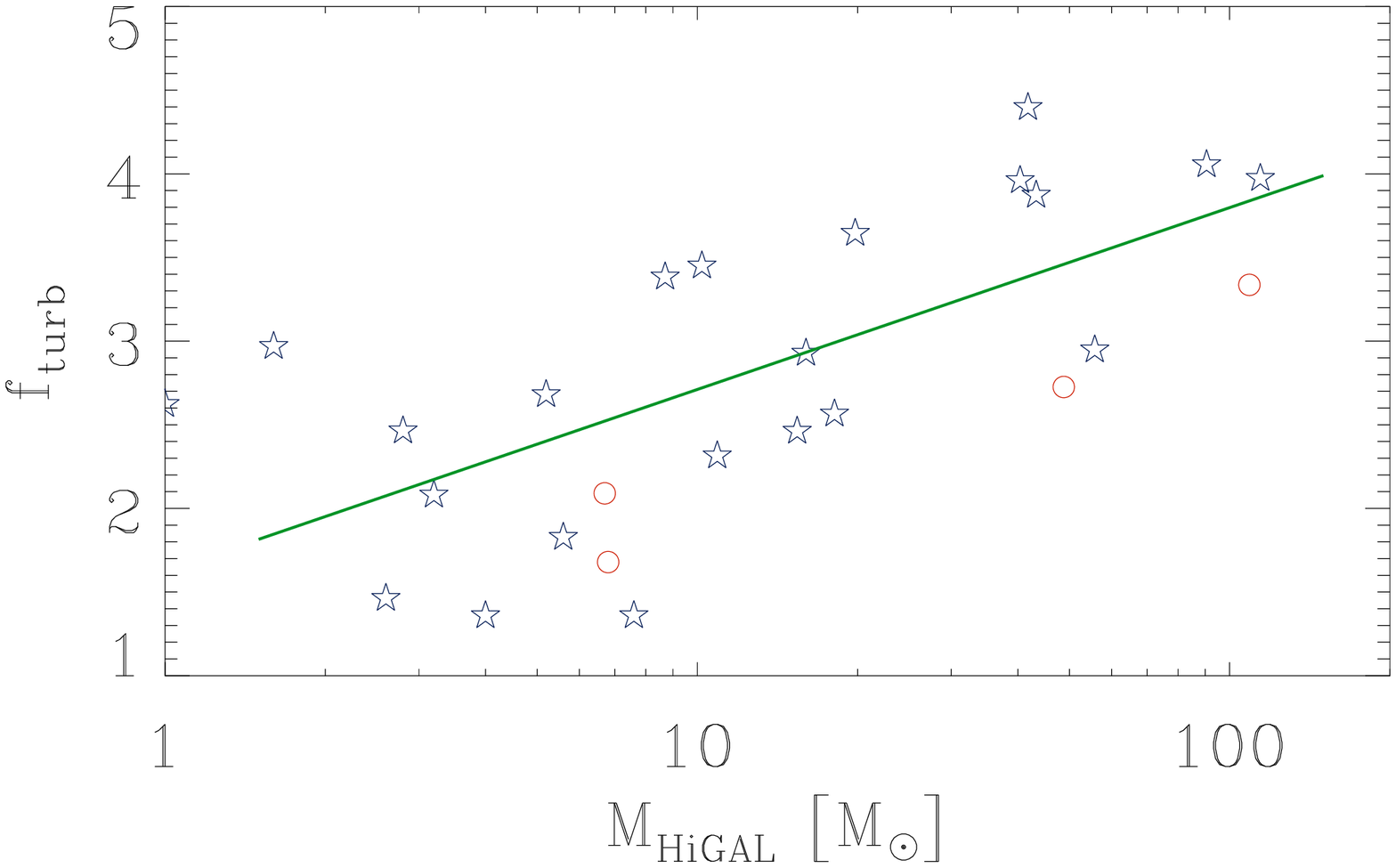}
\caption{
{\it Top panel.} Non-thermal component of the velocity line width of the C$^{18}$O$\,(1-0)$
transition, vs. the total clump mass, $M_{\rm clump} = M_{\rm HiGAL}$ (see text).
Stars represent protostellar clumps and the empty circles represent
starless clumps.
{\it Bottom panel.} Non-thermal velocity dispersion, from C$^{18}$O$\,(1-0)$,
 to sound speed ratio, $f_{\rm turb}$, vs. the total clump mass, $M_{\rm clump}$.
The green solid line represents the linear fit
to all points (see text); the Spearman rank coefficient is 0.66.
}
\label{fig:nonthdisp}
\end{figure}

Another parameter of interest for the analysis of the Hi-GAL clumps is the non-thermal
(turbulent) component of the velocity line widths. If we take a clump's observed  
velocity line width, $\Delta V_{\rm C^{18}O}$, to be the quadrature sum of its thermal 
and non-thermal line widths (assuming that the two contributions are independent of each other), 
$\Delta V_{\rm T}$ and $\Delta V_{\rm NT}$, respectively:
\beq
\Delta V_{\rm C^{18}O}^2 = \Delta V_{\rm T}^2 + \Delta V_{\rm NT}^2 \, ,
\eeq
then in order to find $\Delta V_{\rm NT}$ we must determine $\Delta V_{\rm T}$:
\beq
\Delta V_{\rm T}^2 = 8 \, \log 2 \, \frac{k T}{m_{\rm C^{18}O} } \, ,
\eeq
where $m_{\rm C^{18}O}$ is the mass of the C$^{18}$O molecule. The clump temperature, $T = T_{\rm d}$, 
is taken from the dust-determined Hi-GAL physical parameters. The top panel of
Fig.~\ref{fig:nonthdisp} shows $\Delta V_{\rm NT}$, estimated for all Hi-GAL clumps 
with a reliable C$^{18}$O detection, vs. the total clump mass, 
$M_{\rm clump} = M_{\rm HiGAL}$, also derived from the Hi-GAL data. 
This figure shows a tentative trend of increasing $\Delta V_{\rm NT}$ with $M_{\rm clump}$ and, 
despite the low number of starless clumps in this plot, it also shows that $\Delta V_{\rm NT}$
tends to be higher toward protostellar clumps.  The average values of $\Delta V_{\rm NT}$ are
$1.2\pm 0.3$\,km\,s$^{-1}$  and $1.7\pm 0.6$\,km\,s$^{-1}$ for starless and 
protostellar clumps, respectively.

However, because of the difference in temperature, the physically significant parameter 
is the non-thermal velocity dispersion, $\sigma_{\rm NT}$,
to thermal sound speed, $c_{\rm s}$, ratio which we call $ f_{\rm turb} = \sigma_{\rm NT} / c_{\rm s}$
following \citet{curtis2011}. $c_{\rm s} = \sqrt(\gamma k T / m_{\rm av} )$ is the isothermal
sound speed in the bulk of the gas of mean molecular weight $m_{\rm av}=2.3$~amu (with respect to the 
total number of particles), assuming a mass fraction for He of 25\%. The parameter $\gamma$ is the
adiabatic index ($\simeq 7/5$ for a diatomic molecular gas). In the bottom panel of 
Fig.~\ref{fig:nonthdisp} we then plot $ f_{\rm turb}$ vs. $M_{\rm clump}$, and we note that 
the trend of increasing $ f_{\rm turb}$ with mass is still visible.  We also note that for similar
ranges of mass the $ f_{\rm turb}$ parameter of the starless clumps is typically lower compared
to that of protostellar clumps, though their (full range) average values are $2.46\pm 0.73$ 
and $2.96\pm 1.02$, respectively. In the bottom panel of Fig.~\ref{fig:nonthdisp} we also 
show the result of a linear regression to find the slope of the best fit line, using the
Bayesian IDL routine {\tt LINMIX\_ERR}.

Our results clearly indicate that the line widths toward the Hi-GAL clumps are supersonic.
In addition, since stars can inject significant energy into their surroundings, it is not surprising that
protostellar clumps display somewhat larger non-thermal velocity dispersion compared to starless clumps,
within similar mass ranges. Instead, \citet{curtis2011} do not find a significant difference
between the distributions of $ f_{\rm turb}$ of the starless and protostellar subsets.
Clearly, we need similar samples of starless and protostellar clumps
in order for these differences to be statistically significant. 

\subsection{Derivation of masses }
\label{sec:mass}

\subsubsection{Mass of the filaments derived from column density }
\label{sec:cdmass}


The calculation of the filaments mass 
from the column density assumes LTE and we determine 
the molecular gas mass integrating the molecule column density over a given extent 
of the molecular cloud. Then we can write:

\begin{equation}
M_{\rm cd} = d^2\, m_{\rm mol} \,
\int \, N_{\rm mol} \, {\rm d}\Omega \, ,
\label{Mcd}
\end{equation}
where $\int \, N_{\rm mol} \, {\rm d}\Omega$ is the molecule column
density integrated over the region enclosed by the chosen contour level,
$m_{\rm mol}$ is the mass of the specific molecule being considered, and $d$ is
the distance to the source.
Eq.~(\ref{Mcd}) is actually implemented by writing:
\begin{equation}
M_{\rm cd} = d^2\, m_{\rm mol} \, \Delta \Omega_{\rm pix} \, \sum_{i=1}^{n_{\rm pix}}  
\, N_{\rm mol}(x_i, y_i) \, ,
\end{equation}
where $N_{\rm mol}(x_i, y_i)$ represents the column density in a single pixel $(x_i, y_i)$ of
the map, with $n_{\rm pix}$ representing the total number of pixels, and $\Delta \Omega_{\rm pix}$
represents the solid angle covered by a single pixel. The map pixels selected are those that have an
integrated intensity $I \geq 3\, \sigma_{\rm map}$, 
where the RMS integrated intensity,
$\sigma_{\rm map} \simeq 0.2\,$K\,km\,s$^{-1}$, has been estimated
directly from the map, averaging pixels where no emission is detected.

Using the H$_2$ column density 
derived from the $^{13}$CO and $^{12}$CO$(1-0)$ lines
(shown in the bottom panel of Fig.~\ref{fig:cdC18O}) the 
total gas mass calculated is $\sim 2.5 \times 10^3 \,$M$_\odot$, including the isolated tile
shown in Fig.~\ref{fig:map250}, using the relative molecular
abundances described in Section~\ref{sec:coldens}. Specifically, we obtain 
$\simeq 1.8 \times 10^3 \,$M$_\odot$ and $\simeq 6.0 \times 10^2 \,$M$_\odot$ for the
main and secondary filaments, respectively.
%
%
For comparison, if we use the H$_2$ column density derived from the $^{13}$CO and C$^{18}$O$(1-0)$ lines
(shown in the top panel of Fig.~\ref{fig:cdC18O}) the 
total gas mass calculated would be $\sim 4.4 \times 10^3 \,$M$_\odot$.


\subsubsection{Clumps virial mass }
\label{sec:virial}


%
\begin{figure}[!ht]
\centering
\includegraphics[width=9.5cm,angle=0]{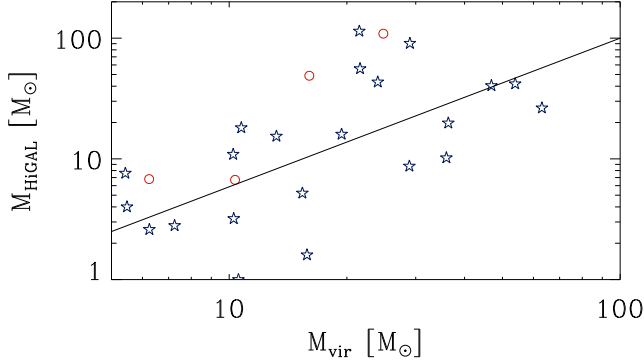}
\caption{
Total clump mass, $M_{\rm clump} = M_{\rm HiGAL}$,  vs. the virial mass, $M_{\rm vir}$, calculated
using the velocity line widths of the C$^{18}$O$\,(1-0)$ transition  (see text).
The solid line indicates the minimum $M_{\rm clump} = 0.5 M_{\rm vir}$ for
which the clump should be self-gravitating. Symbols are as in Fig.~\ref{fig:nonthdisp}.
}
\label{fig:virial}
\end{figure}

The question of whether molecular clumps and their sub-structures are bound is of 
great importance in understanding star and cluster formation.
The virial mass estimate, in combination with other mass estimates, may allow 
to determine whether or not a specific clump is bound and virialized. 
We estimate the virial mass of the clumps assuming they are simple spherical systems
with uniform density \citep{maclaren1988}: 
%
\begin{equation}
M_{\rm vir}{\rm [M_\odot]} = 210 \, R_{\rm dec}{\rm [pc]} \, (\Delta V_{\rm av} {\rm [km~s^{-1}]})^2 \, ,
\label{eq:mvir}
\end{equation}
where $R_{\rm dec}$ is the deconvolved source radius and $\Delta V_{\rm av}$ represents 
the line FWHM of the molecule of mean mass, calculated as the sum of the 
thermal and turbulent components (see Section~\ref{sec:linewidths}):
%
%
\begin{equation}
\Delta V_{\rm av}^2 = \Delta V_{\rm mol}^2 + kT8 \ln 2 \, \left ( \frac{1}{m_{\rm av}} -
\frac{1}{m_{\rm mol}} \right ) \, ,
\end{equation}
where $\Delta V_{\rm mol}$ is the line FWHM of the molecular transition being considered.
We note that according to \citet{maclaren1988}, Eq.~(\ref{eq:mvir}) may lead
to a mass overestimate if the density distribution is not uniform. For example, in the
case of a sphere with a density distribution $\rho \propto r^{-2}$ the numerical factor in 
Eq.~(\ref{eq:mvir}) should be replaced by 126.

As described in Section~\ref{sec:linewidths}
the line FWHM is calculated from the C$^{18}$O 
spectrum observed at the nominal position of each Hi-GAL clump. 
Since even C$^{18}$O may not trace the denser molecular gas and may be affected by the
ambient, less dense material, we have corrected its line width for the average ratio 
of the C$^{17}$O to C$^{18}$O line width ratio, which is 0.84 toward 
the bright region of emission
in the NW section of the main filament (shown as a black box in Fig.~\ref{fig:maps}) 
In addition, we have used the numerical factor 126 in Eq.~(\ref{eq:mvir}), 
since for the density profiles of molecular clumps power-laws
have been put forward by various studies (see, e.g., \citealp{fontani2002} where they 
find $n \sim r^p$ with an average $p=-2.59$).

The plot of the total 
clump mass, $M_{\rm HiGAL}$, as derived from Hi-GAL, vs. the virial mass is shown in 
Fig.~\ref{fig:virial}.  We note that all of the starless clumps where the C$^{18}$O line
has been detected lie above the self-gravitating line, indicating that they are likely to be 
gravitationally bound. The protostellar clumps, much more numerous, are almost equally divided 
between those above and below the self-gravitating line.
In interpreting this plot, however, one should consider that the line widths we measure are 
from a region (determined by the Mopra telescope beam) that is generally different than the 
virial radius we assign to it, which is
instead determined from the source-extraction procedure used for the Hi-GAL maps. 
Whether the continuum emission
is more extended or not than the dense molecular gas emitting volume depends on the specific 
excitation conditions of the gas and the molecular tracer.    
%
%
Therefore, given the uncertainties in both the virial and Hi-GAL dust mass estimates, it is not possible 
to draw unequivocal conclusions about the stability of individual clumps.  

\subsubsection{Dust continuum masses }
\label{sec:dustmass}


Obtaining mass estimates of molecular clouds with molecular lines may be problematic
due to high column densities and also molecular freezeout onto dust grains in some cases.
Alternatively, the cold dust emission, as observed in the (sub)millimeter, can be used to
provide more reliable mass estimates as it is optically thin and does not deplete.

We use the $250\,\mu$m dust continuum emission from Hi-GAL, to achieve isothermal mass
estimates using the expression:
\beq
M_{\rm cont} = \frac{I_{\rm \nu} \, d^2}  {r \, k_{\rm 250} \, B_{\rm \nu}(T_{\rm d}) } \, ,
\eeq
where $I_{\rm \nu}$ is the source surface brightness (or specific intensity) in the Hi-GAL map at
$250\,\mu$m (MJy\,sr$^{-1}$), $d$ is the distance (cm),
$r\, k_{\rm 250}$  is the dust mass absorption coefficient (cm$^2$\,g$^{-1}$), often called the opacity,
corrected for the dust-to-gas ratio, $r$, at $\lambda_\circ = 250\,\mu$m,
and $B_{\rm \nu}(T_{\rm d})$ is the Planck function at dust temperature, $T_{\rm d}$.
We adopt a value of $r\, k_{\rm 250} = 0.11$\,cm$^2$\,g$^{-1}$ from \citet{martin2012}.

We integrate $I_{\rm \nu}$ in the same area as mapped at Mopra, as shown in Fig.~\ref{fig:map250},
and also assume that $T_{\rm d}$ can vary in the same range analyzed by \citet{martin2012}, that is,
$\simeq 12 - 17$\,K. We then obtain $M_{\rm cont} \sim 2 \times 10^3$\,M$_\odot$ ($T_{\rm d} = 12$\,K)
and $M_{\rm cont} \sim 5 \times 10^2$\,M$_\odot$ ($T_{\rm d} = 17$\,K).
If we use a typical intermediate value of $T_{\rm d} = 14$\,K, the total mass is
$M_{\rm cont} \sim 10^3$\,M$_\odot$ and specifically, $M_{\rm cont} \sim 7.6 \times 10^2$ and
$\sim 1.9 \times 10^2$\,M$_\odot$ for the main and secondary filaments, respectively.
Given all uncertainties in both molecular abundances and dust parameters, the dust continuum
masses agree reasonably well with the column density masses derived in Section~\ref{sec:cdmass}.
Also the main to secondary filament mass ratios estimated from molecular lines and dust
continuum are very similar ($M_{\rm main}/M_{\rm sec} \simeq 3-4$), showing that the
secondary filament has about 1/3 of the mass of the main filament. The good agreement of
the mass ratio, $M_{\rm main}/M_{\rm sec}$, derived with these two methods, 
suggests that the difference between the absolute mass values, $M_{\rm main}$ and $M_{\rm sec}$, derived
from the molecular lines and the dust continuum depends on the choice of parameters such as
molecular abundances and dust mass absorption coefficient.

%
%
%
\begin{figure*}[!ht]
\centering
\includegraphics[width=9.0cm,angle=0]{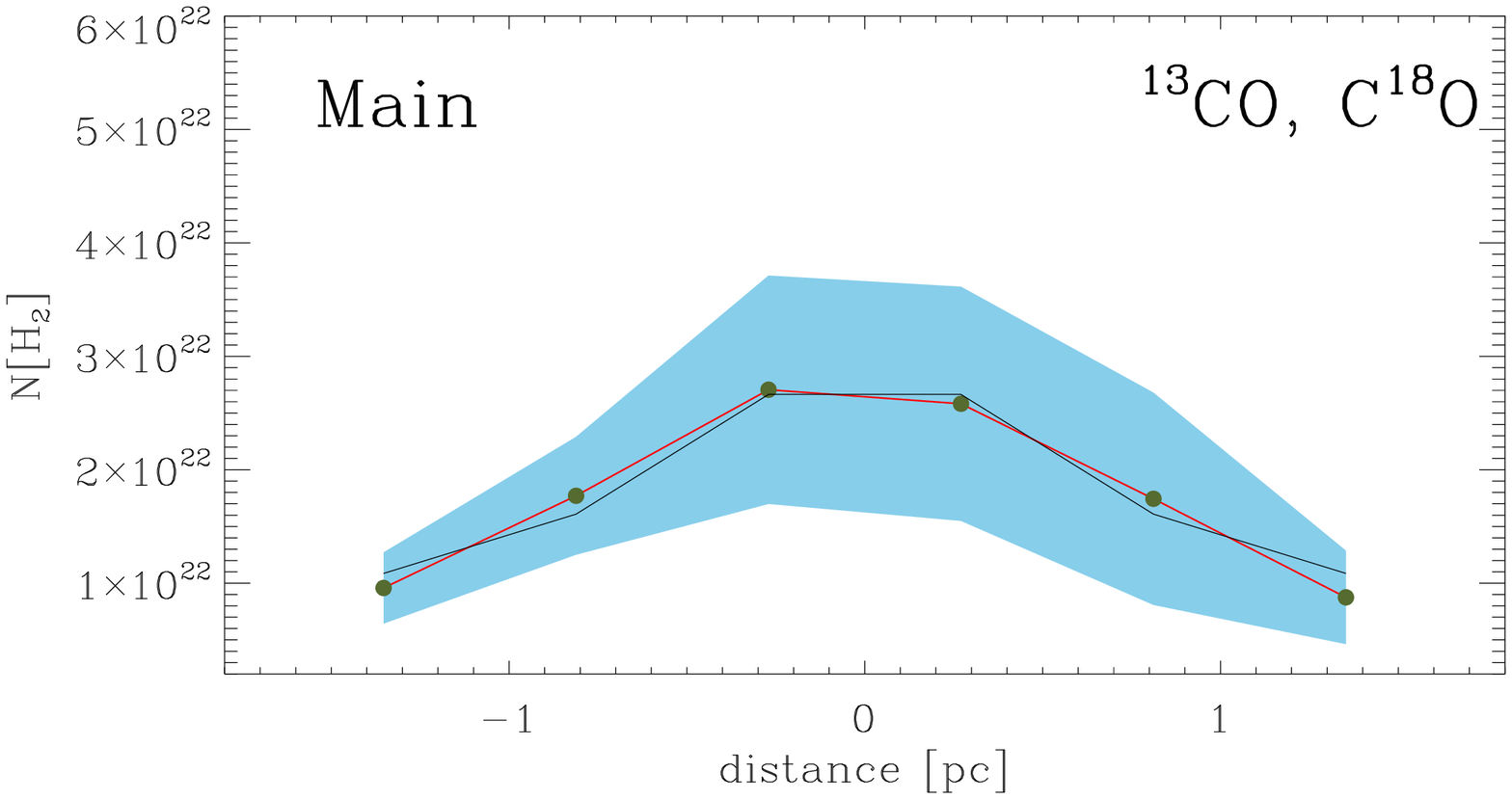}
\includegraphics[width=9.0cm,angle=0]{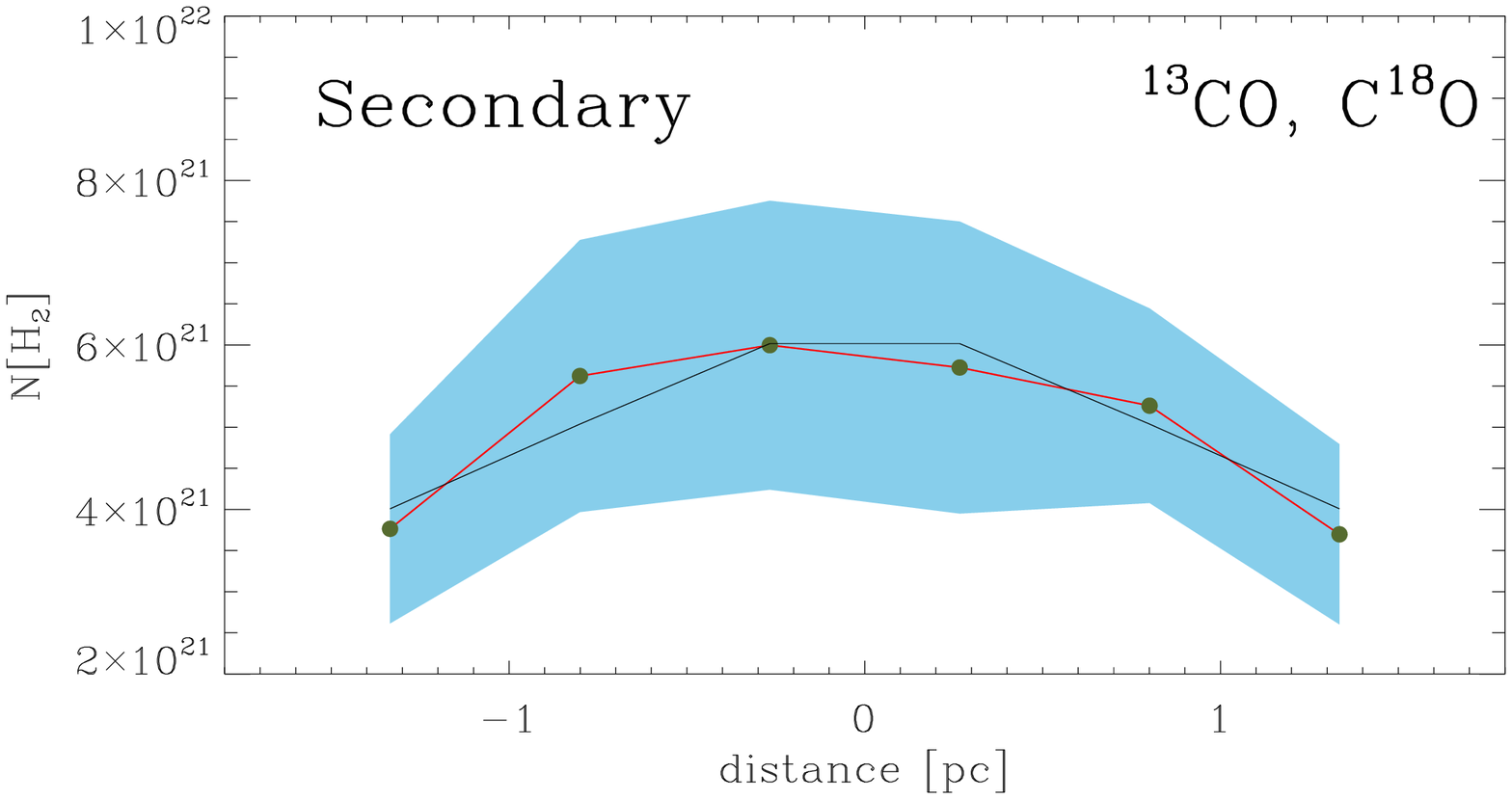}
%
%
\includegraphics[width=9.0cm,angle=0]{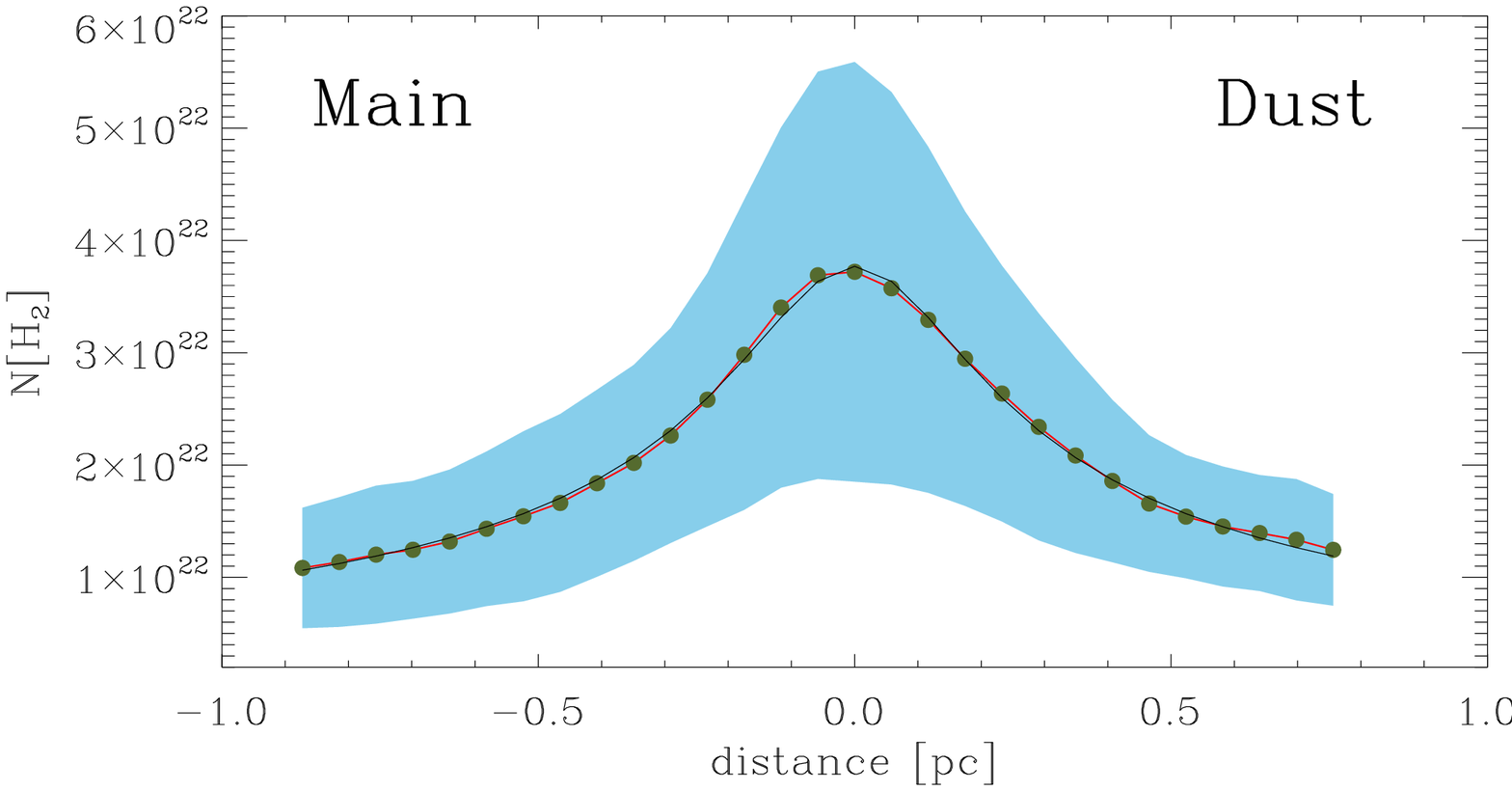}
\includegraphics[width=9.0cm,angle=0]{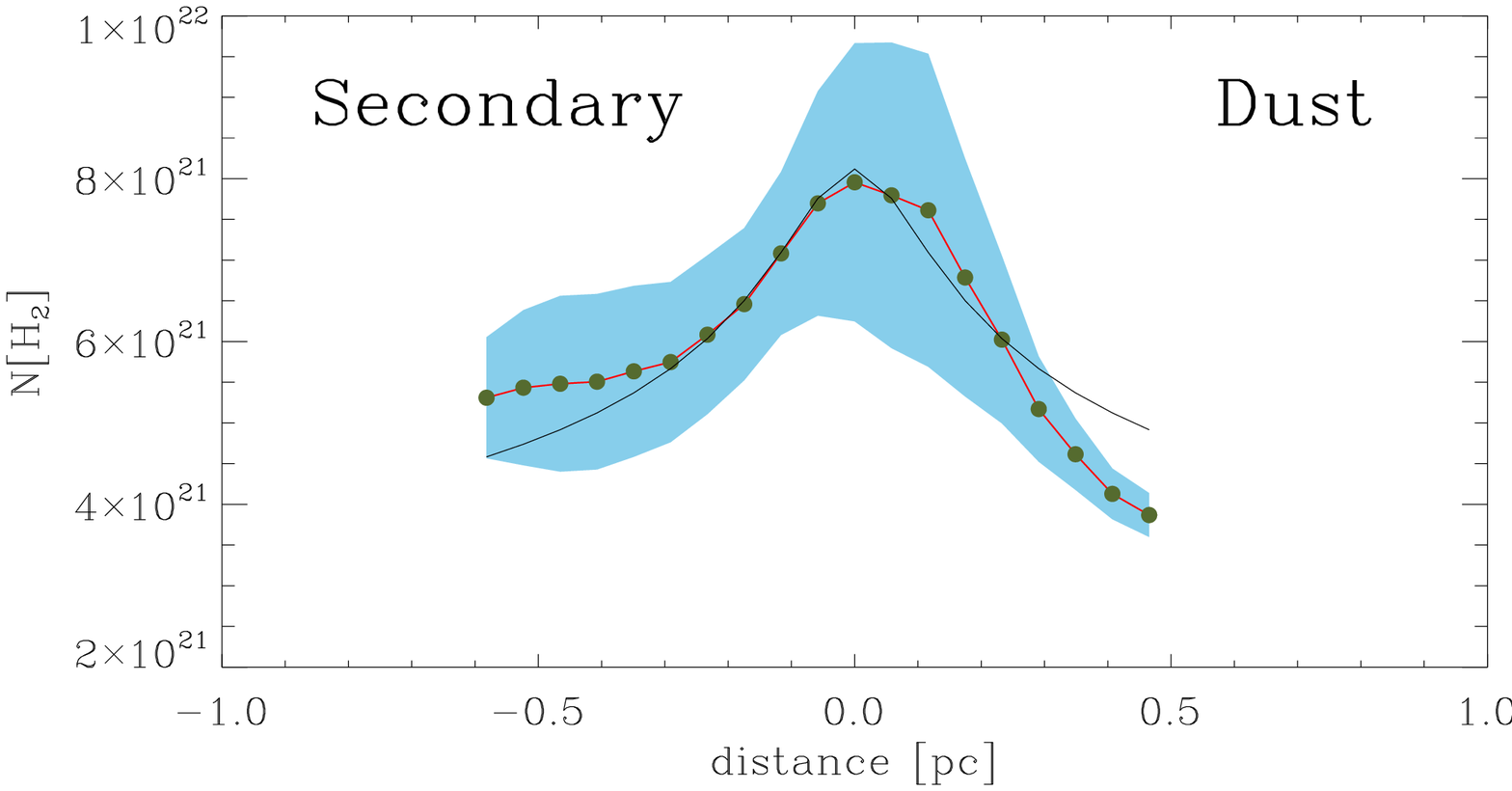}
\caption{
Radial column density profile of the main (left column) and secondary (right column)
filaments. The dots, connected by the red solid line, show the mean profile across the filaments, while the
shaded area indicates the standard deviation. In the top row the average profiles
were estimated from the $^{13}$CO and C$^{18}$O$(1-0)$ derived column density (Fig.~\ref{fig:cdC18O},
top panel), while in the bottom row we used the column density estimated by \citet{elia2013}.  
The black solid line represents the best fit profile, following
the formulation in \citet{arzoumanian2011}. 
}
\label{fig:cdprofile}
\end{figure*}

\section{Discussion}
\label{sec:discussion}

The results presented in Sections~\ref{sec:results} and \ref{sec:analysis} can be used to discuss
the physical properties and evolutionary state of the two filaments.

\subsection{Column density profiles}
\label{sec:cdprof}



In this section, we analyze the radial column density profile of the filaments
using both molecular line emission and dust submillimeter continuum emission.
To construct the mean radial density profile of each filament we used the following procedure.
First, we searched for the location of peak emission along every cut in declination through each filament. 
A simple scanning in declination allowed us to unambiguously find the peak emission because the 
main axis of each filament make a large ($> 45^\circ$) angle 
to the declination axis. Second, we constructed the midline of each filament by joining all peak ridge points. 
We then measured a
radial column density profile perpendicular to the midline at each position. Finally, we derived the
mean radial profile by averaging all profiles along each filament. The mean profiles and standard
deviations, at each radial separation, are shown in Fig.~\ref{fig:cdprofile}.

In order to characterize each observed column density profile, following \citet{arzoumanian2011}
we adopt the model of a cylindrical filament with radial column density profile (as a function of
cylindrical radius $r$) of the form:
\beq
N_{{\rm H}_2}(r) = A_{\rm p} \, \frac{ \rho_{\rm c} R_{\rm flat} }    { [1 + (r/R_{\rm flat})^2 ]^\frac{p-1}{2} } \, ,
\label{eq:cdprofile}
\eeq
where $N_{{\rm H}_2}(r)$ is the column density at radial separation $r$, $\rho_{\rm c}$ is the
number density at the center of the filament, $R_{\rm flat}$ is the characteristic radius of the
flat inner portion of the density profile, and the parameter $ A_{\rm p}$, for $p>1$, 
is calculated as:
\beq
 A_{\rm p} = \frac{ 1} {\cos i } \int_{-\infty}^\infty \frac { {\rm d} u}  {(1 + u^2)^{p/2} }  \, ,
\label{eq:Ap}
\eeq
where the inclination angle along the line of sight, $i$, is assumed to be 0 for simplicity.
Allowing all three parameters, $p$,  $\rho_{\rm c}$ and $R_{\rm flat}$  
to vary, we find the values listed in Table~\ref{Table:cdprofile}, 
while the best fit profiles are shown in Fig.~\ref{fig:cdprofile}.
For the dust continuum emission we used the column density map derived by \citet{elia2013} for
better accuracy. In fact, they obtained the column density after performing a fit to the
SED at each point in the Hi-GAL maps, thus deriving the dust temperature at the same time.

%
%
\begin{table*} [!ht]
\centering
\caption{Parameters corresponding to the best-fit model of the form expressed by Eq.~(\ref{eq:cdprofile})
to the mean column density profile of the main and secondary filaments. 
}
\begin{tabular} {lccccccc}
\hline\hline
\noalign{\smallskip}
Filament     & \multicolumn{3}{c}{\bf Main filament}   &   & \multicolumn{3}{c}{\bf Secondary filament}  \\ 
\cline{2-4}
\cline{6-8}
         & $\rho_{\rm c}$            & $R_{\rm flat}$         & $p$    &
         & $\rho_{\rm c}$            & $R_{\rm flat}$         & $p$    \\
\noalign{\smallskip}
         & [$\times 10^5$ cm$^{-3}$]   & [pc]   &         &          & [$\times 10^5$ cm$^{-3}$]   & [pc]  &  \\
\hline
\noalign{\smallskip}
Spectral lines         & 0.6    & 0.48    & 1.9   &   & $-$   & $-$     & $-$  \\
Dust emission   & 1.8    & 0.19    & 1.8   &   & 0.3   & 0.10    & 1.3    \\
\noalign{\smallskip}
\hline

\label{Table:cdprofile}
\end{tabular}

\end{table*}
%

Comparing the best-fit values of Table~\ref{Table:cdprofile} with those in the literature, 
we find that the values of $p$ are similar to those found by
\citet{arzoumanian2011} and \citet{andre2016} (who used dust derived column density maps),  
and also by \citet{kirk2013} (who used molecular lines). However, \citet{nutter2008} found
a value of $p$ somewhat higher ($p=3$)  towards TMC1.
Instead, the best fit values of the other two free parameters, $\rho_{\rm c}$ and $R_{\rm flat}$, 
significantly differ from those of \citet{kirk2013},  
\citet{arzoumanian2011} (who do not list their $\rho_{\rm c}$ values) and \citet{andre2016},
particularly when the values derived from the molecular lines are considered. 
However, our best-fit values for all three parameters
are quite similar to those found by \citet{juvela2012b}, who in addition analyzed sources
with a range of distances more similar to that of the $\ell=224^{\circ}$ region, and hence
with similar distance-related effects.

The differences found in the literature are certainly due, at least in part,
on the way the analyzed structures are selected and on the analysis methods themselves 
(see for example the discussion in \citealp{juvela2012b}).
Specifically, although both of the filaments studied here have been identified by \citet{schisano2014},
they are not as well defined (e.g., with a sharper central 
peak in the column density average profile) 
as those mapped by \citet{kirk2013} and \citet{arzoumanian2011},
which also lie at quite a different distance, 
or also simulated (see, e.g., \citealp{juvela2012a}). 
It is not thus entirely surprising that we obtain
lower values of the filament central density and a higher value of $R_{\rm flat}$.


In Table~\ref{Table:cdprofile} and Fig.~\ref{fig:cdprofile}, 
comparing the column density profiles obtained from the molecular lines and from the dust
continuum emission, the observed discrepancies are likely caused in part by the different radial extent 
measurable in the Mopra and Hi-GAL maps. In fact, when using the column density
derived from molecular line emission the reliability of our fit is limited by the fact that
we were able to measure the radial column density profile only relatively close to the central ridge.
For the specific case of the secondary filament, the radial column density profile from the Mopra
data is more uncertain due to the lower SNR in this region of the map, particularly at the edges
of the filament, and thus the best-fit parameters are not shown in Table~\ref{Table:cdprofile}.
The main filament should be less affected by SNR problems and the different
radial profiles shown in Fig.~\ref{fig:cdprofile} must thus be caused also by the
method used to retrieve the column density and/or the intrinsic distribution of the material
detected with each technique.
The shallow or flat radial column density profile derived from the Mopra spectral line data
suggests that we are probably sensitive to some amount of large-scale structures, that is,
diffuse material not directly associated with the star-forming regions along the filament,
which is more reliably traced by the optically thin dust emission.








\subsection{Gravitational stability of the filaments}
\label{sec:gravstab}

The properties of interstellar filaments and possible scenarios for their
formation and evolution have already been discussed in many papers, both
theoretically (e.g., \citealp{ostriker1964}, \citealp{inutsuka1997}, \citealp{inutsuka2001}, \citealp{fiege2000})
and observationally (e.g., \citealp{andre2010}, \citealp{arzoumanian2011,arzoumanian2013},
\citealp{kirk2013}). A proposed scenario, especially supported by the results of the
HGBS, is that large-scale turbulence,
rather than large-scale gravity, plays the dominant role in forming interstellar filaments.
However, gravity would appear to be the major driver in the subsequent evolution of the filaments,
which may contract and accrete material from the background cloud, thus causing filaments to fragment
into clumps and form (proto)stars under certain physical conditions.

Observationally, the gravitational fragmentation of filaments can be analyzed by comparing
our Hi-GAL maps to existing models of filamentary cloud fragmentation. \citet{inutsuka1997}
(and references therein) showed that an unmagnetized isothermal filament is unstable to axisymmetric
perturbations if the line mass or mass per unit length, $M_{\rm line}$, of
the filament is larger than the critical value required for equilibrium,
$M^{\rm unmag}_{\rm line,crit} = 2\,c_s^2/G$, where
$G$ is the gravitational constant.  As shown by \citet{ostriker1964}, the critical line mass
depends on gas temperature only and, following \citet{kirk2013}, $M^{\rm unmag}_{\rm line,crit}$
can also be written in more convenient units as:
\beq
M^{\rm unmag}_{\rm line,crit} = 16.7 \, \left ( \frac{T } {10{\rm K} } \right ) \, \, M_\odot \, {\rm pc}^{-1} \, ,
\eeq
therefore if $T$ is in the range of $\simeq 10$ to 15\,K, then $M^{\rm unmag}_{\rm line,crit}$
is between $\simeq 17$ and 25\,$M_\odot \, {\rm pc}^{-1}$.
To derive the filament masses per unit length from our Hi-GAL maps, we used
the radial column density profiles discussed in Section~\ref{sec:cdprof}.
The observed mass per unit length, $M_{\rm line}$, of the main and secondary filaments
was then derived by integrating the measured average column density profile
(as shown in Fig.~\ref{fig:cdprofile}) over radius, i.e.,
$M_{\rm line} = \mu_{{\rm H}_2} \, m_{\rm H} \, \int \langle N_{{\rm H}_2}(r) \rangle \, {\rm d}r$, 
where $\mu_{{\rm H}_2} = 2.8$ is the molecular weight per hydrogen molecule \citep{kauffmann2008} and 
$m_{\rm H}$ is the H-atom mass.

Because it is difficult to exactly define the edges of a filament, we decided to
limit the range of integration to the 50\% level of the peak value.
If we use the $250\,\mu$m radial column density profile we obtain
$M_{\rm line} \simeq 230$ and $\simeq 50$\,$M_\odot \, {\rm pc}^{-1}$
for the main and secondary filaments\footnote{No projection effects have been considered.},
respectively. Both filaments would then appear to be
supercritical, that is, gravitationally unstable with $M_{\rm line} > M^{\rm unmag}_{\rm line,crit}$.
As a comparison, the values of $M_{\rm line}$ found in the literature vary in a wide range, from a few 
$M_\odot \, {\rm pc}^{-1}$ (e.g., \citealp{arzoumanian2013}, \citealp{juvela2012b} and \citealp{schisano2014}) to as much as
several 100s of $M_\odot \, {\rm pc}^{-1}$ (e.g., \citealp{contreras2013}, \citealp{li2016} and \citealp{schisano2014}).

We note that while the main filament is highly supercritical, with $M_{\rm line}$ roughly
ten times larger than the critical value, the secondary filament has a much lower $M_{\rm line}$ value.
The total amount of support available in the main filament is therefore likely insufficient
to keep the filament in equilibrium, implying that it should be radially contracting.
This scenario assumes that unstable filaments
would accrete additional mass from their surroundings and increase in mass per unit length
while contracting with time.  At the same time as they contract and accrete material from
the background cloud, supercritical filaments are expected to fragment into clumps and form
(proto)stars, giving rise to the morphology of the main filament and clumps  shown in
Fig.~\ref{fig:map250}.
However, as we noted in Section~\ref{sec:linewidths}, we find no observational evidence
showing that the gas is globally and radially  infalling in either filament.
This is either due to observational effects, related to the spectral lines used,
or the radially contracting phase has already slowed down or stopped
(or possibly not started yet, in the case of the secondary filament), thus being
effectively undetectable.

Another alternative to explain the non-detection of radial infall toward the main filament
is that all of the non-thermal motion is contributing thermal-like support.
In this case one must consider the virial mass per unit length, $M_{\rm line,vir} = 2\,\sigma_{\rm av}^2/G$
\citep{fiege2000}, where $\sigma_{\rm av}$ is the velocity dispersion of the molecule
of mean mass (see Section~\ref{sec:virial}):
\beq
\sigma_{\rm av} = \frac{ \Delta V_{\rm av} } {\sqrt{8 \ln 2 } } \, .
\eeq
%
%
We note that $M_{\rm line,vir}$ is analogous to the the virial mass and thus the dividing value
between gravitationally unbound  and bound filaments is defined by $M_{\rm line,vir} = 2 \, M_{\rm line}$.
Figures~\ref{fig:kinematicsC18O} and \ref{fig:kinematicsCONV} can be used to estimate a
typical velocity dispersion for the main filament, $\sigma_{\rm av} \sim 0.6\,$km\,s$^{-1}$, and the
secondary filament, $\sigma_{\rm av} \sim 0.4\,$km\,s$^{-1}$. Thus we estimate $M_{\rm line,vir}$ $\sim 70$
and $\sim 160$\,$M_\odot \, {\rm pc}^{-1}$ for the main and secondary filaments, respectively.
Therefore, the thermally supercritical main filament is found to be a self-gravitating structure.
The dynamical state of the secondary filament is instead much less defined, due to the uncertainty
in the actual value used for $\sigma_{\rm av}$, and is thus at the edge of being in
virial equilibrium or gravitationally unbound.

\subsection{A possible evolutionary scenario}
\label{sec:evolution}

The noticeable separation between the (less evolved, younger; see Fig.~\ref{fig:LvsM}) starless clumps and the
(more evolved, older) protostellar clumps on the secondary and main filament, respectively, 
and the fact that the more massive and highly fragmented  main filament
is thermally supercritical and gravitationally bound, suggest a later stage of evolution
of the main filament. In the main filament clumps have accreted enough mass to become
mostly protostellar, and the molecular gas surrounding them appears to be more turbulent compared
to starless clumps, as shown in Section~\ref{sec:linewidths}. Also, the more diffuse gas in the
main filament shows a higher degree of turbulence compared to the secondary filament. A trend
of increasing $\sigma_{\rm av}$ with $M_{\rm line}$ has in fact been discussed by
\citealp{arzoumanian2013}. 

As mentioned in Section~\ref{sec:velgrad} we observe a velocity
gradient along the main axis of the filament,
and we have discussed several possible interpretations for this gradient.
It is however interesting to discuss in greater details the explanation 
of this velocity gradient in terms of a possible accretion flow from the filament 
onto the central cluster.
A similar scenario has indeed been proposed for the case of the Serpens South
filament and cluster, though with a higher value of ${\rm d}V/{\rm d}r$ \citep{kirk2013}.
A difficulty with this interpretation is that in our main filament the protostellar clumps are 
actually distributed along the entire length
of the structure  (see Figures~\ref{fig:map250} and  \ref{fig:kinematics13CO}), though there is
a main cluster of clumps in the NW section of the filament.

The much less massive secondary filament, which is also marginally supercritical, appears
to have a smoother and regular profile (as confirmed by the narrow shaded area, indicating
the standard deviation, in Fig.~\ref{fig:cdprofile}). As we mentioned already 
it contains only starless clumps, and the less turbulent ambient gas and 
its dynamical state (gravitationally unbound or near to
virial equilibrium) suggest that this structure could be in an earlier evolutionary phase.
Alternatively, this low-density filament may have evolved differently from the main filament, and its
current dynamical state may prevent any further accretion.
However, our spectral line maps are not extended enough to determine whether molecular gas is 
actually flowing from the ambient gas also toward the secondary filament, and whether it
could eventually follow the same evolutionary path of its main counterpart.

If this low-density filament is indeed unbound it may be expected
to disperse on a turbulent crossing time, unless it is confined by some external pressure.
\citet{arzoumanian2013} also suggest that low-density filaments, observed roughly
perpendicular to the main supercritical filaments, could be feeding them contributing to their
accretion process and growth in mass per unit length.
The velocity gradient measured along the secondary filament and shown in 
Fig.~\ref{fig:kinematics13CO} is roughly directed toward the main filament, and appears
to point toward the bright region
located at the center-south (RA\,$\sim 07^{\rm h} \, 10^{\rm m}$, DEC\,$\sim -10^\circ \, 30'$)
of both Hi-GAL and spectral line maps (see Figures~\ref{fig:map250}, \ref{fig:maps} and
\ref{fig:cdC18O}), which also corresponds to a higher concentration of protostellar clumps.
However, as noted in Section~\ref{sec:velgrad}, this velocity gradient is less 
reliable compared to that measured along the main filament. 

Therefore, if we use the simple method discussed by \citet{kirk2013} 
to estimate the mass accretion rate implied by the velocity gradient along the main filament, we 
obtain an accretion rate along the filament onto the two main clusters of protostellar 
clumps of $\sim100 \,M_\odot/$Myr, assuming the  inclination of the filament 
to the plane of the sky is exactly 0, and also assuming that the observed velocity gradient has not
an entirely different kinematical origin (e.g., rotation).
We can compare this estimate of the accretion rate with the star-formation rate (SFR),
$\simeq260 \,M_\odot/$Myr, found by \citet{elia2013} for the $\ell=224^{\circ}$ region. 
Thus we can see that our estimate of the mass accretion rate is comparable with the mass 
supply required to form the current generation of YSOs along the main filament.
This result suggests that filamentary accretion may significantly affect the formation 
and evolution of stars along the main filament.

%
%

\section{Conclusions}
\label{sec:conclusions}

Using the Hi-GAL dust continuum information and analyzing the Mopra spectral line maps, 
we study the distribution of starless and protostellar clumps as well as the dynamics 
of the two filaments observed toward the $\ell=224^{\circ}$ region.
The protostellar clumps are more luminous and more turbulent compared to the starless clumps, 
and also lie in regions where the filamentary ambient gas shows larger linewidths. 
The more massive and highly fragmented main filament appears to be thermally supercritical 
and gravitationally bound, suggesting a later stage of evolution compared to the 
secondary filament, which is gravitationally unbound (or near to virial equilibrium) and 
hosts less turbulent ambient gas. Therefore, the low-density secondary filament is either
in an earlier evolutionary phase, or it may have evolved differently from the main filament, and its
current dynamical state may prevent any further accretion.
The status and evolutionary phase of the
Hi-GAL clumps would then appear to correlate with that of the host filament.

Our observations show a velocity gradient along the main filament. If this gradient 
is indeed associated with a mass flow along the filament, possibly feeding the main 
two clusters of protostellar objects, then the estimated mass accretion rate would be 
high enough to lead to the formation of several tens of YSOs
(of mean mass 0.5\,M$\odot$) in a few Myr.
Our observations also indicate that there might be gas flow from the secondary filament 
onto the two main clusters of protostellar clumps located on the main filament. 
To assess the reliability of this tentative scenario as well as
to investigate the presence of infalling gas from the filament onto the clusters,  the area
covered by our Mopra maps would need to be extended and different spectral lines,
such as HCO$^+$ and H$^{13}$CO$^+ \, (1-0)$, should be used.

%


\begin{acknowledgements}
The Mopra radio telescope is part of the Australia Telescope National
Facility which is funded by the Australian Government for operation as a
National Facility managed by CSIRO.
The University of New South Wales Digital Filter Bank used for the
observations with the Mopra Telescope was provided with support from
the Australian Research Council.
The authors wish to thank the staff of the Mopra Observatory for the support provided
before and during the observations. D.E.'s research activity is supported by the
European Union Seventh Framework Programme (FP7/2007-2013) under grant
agreement no. 607380 (VIALACTEA: the Milky Way as a star-formation engine).
\end{acknowledgements}

\bibliographystyle{aa}
\bibliography{refs}

\appendix

\clearpage

\section{Channel maps }

%
%
%
%
%
\begin{figure*}
\centering
\hspace*{-0.5cm}
\includegraphics[width=18.0cm,angle=270]{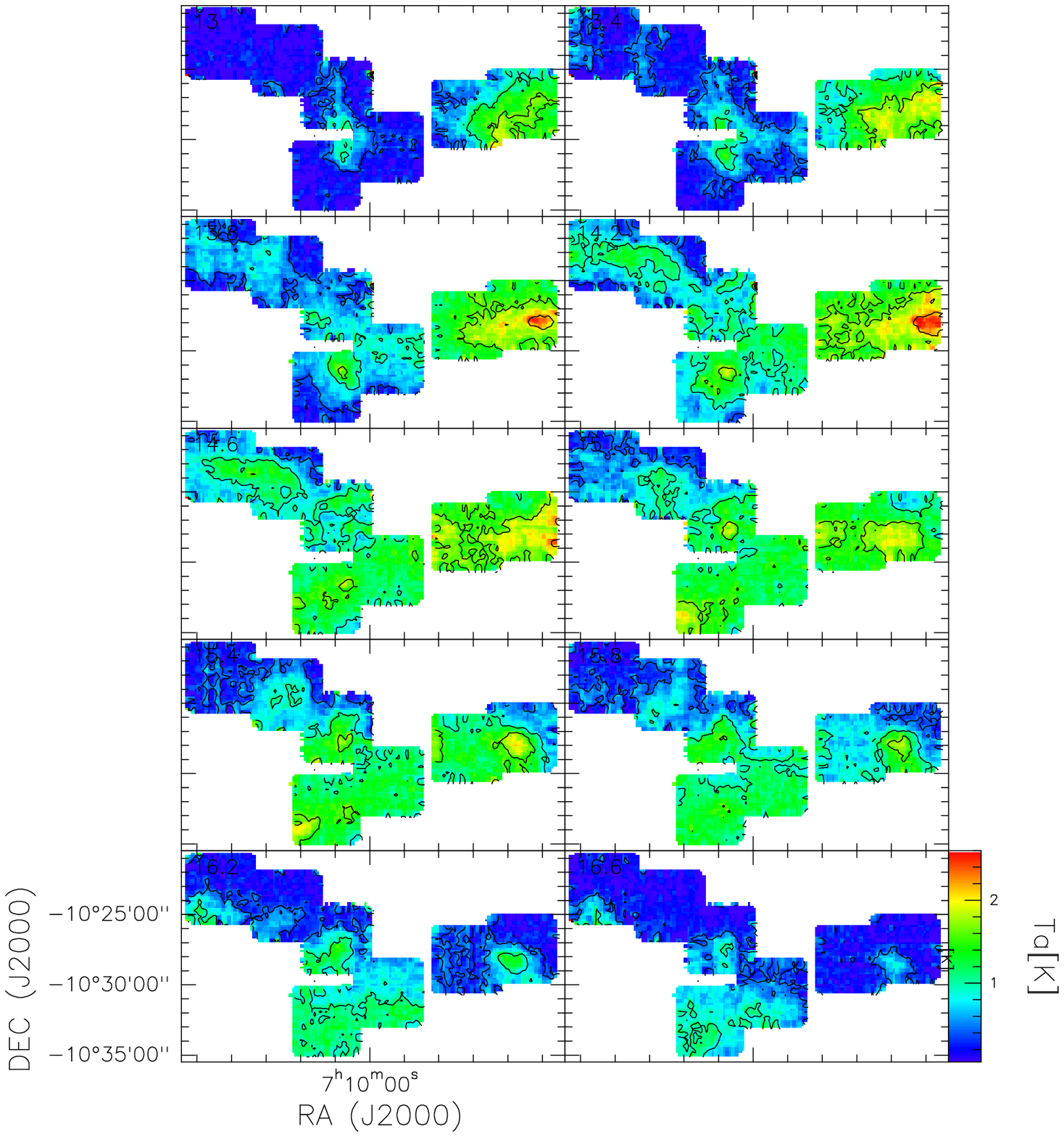}     
\caption{
Channel maps of $^{13}$CO$(1-0)$, between 13 and 16.6 km\,s$^{-1}$ (as shown in the top-left
corner of each panel). The isolated
individual region, centered at (RA,DEC)$\sim (07:11:18, -10:36:00)$, has been
dropped from this figure to increase visibility of the main regions of emission.
}
\label{fig:chmap13CO}
\end{figure*}

%
%
%
\begin{figure*}
\centering
\hspace*{-0.5cm}
\includegraphics[width=18.0cm,angle=270]{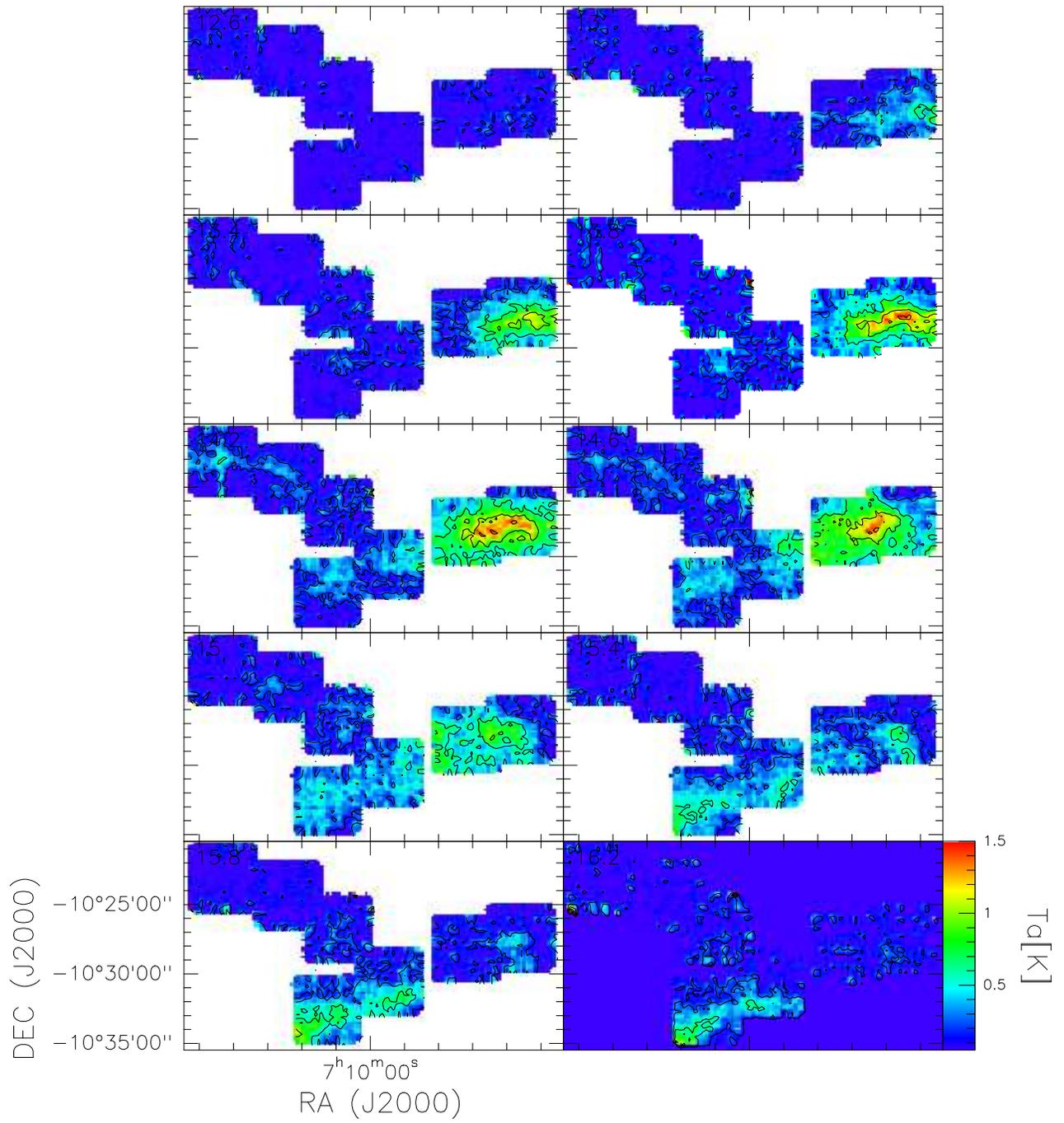}   
\caption{
Same as Fig.\ref{fig:chmap13CO} for C$^{18}$O$(1-0)$.
}
\label{fig:chmapC18O}
\end{figure*}

%
%
%
\begin{figure*}
\centering
\hspace*{-0.5cm}
%
\includegraphics[width=18.0cm,angle=270]{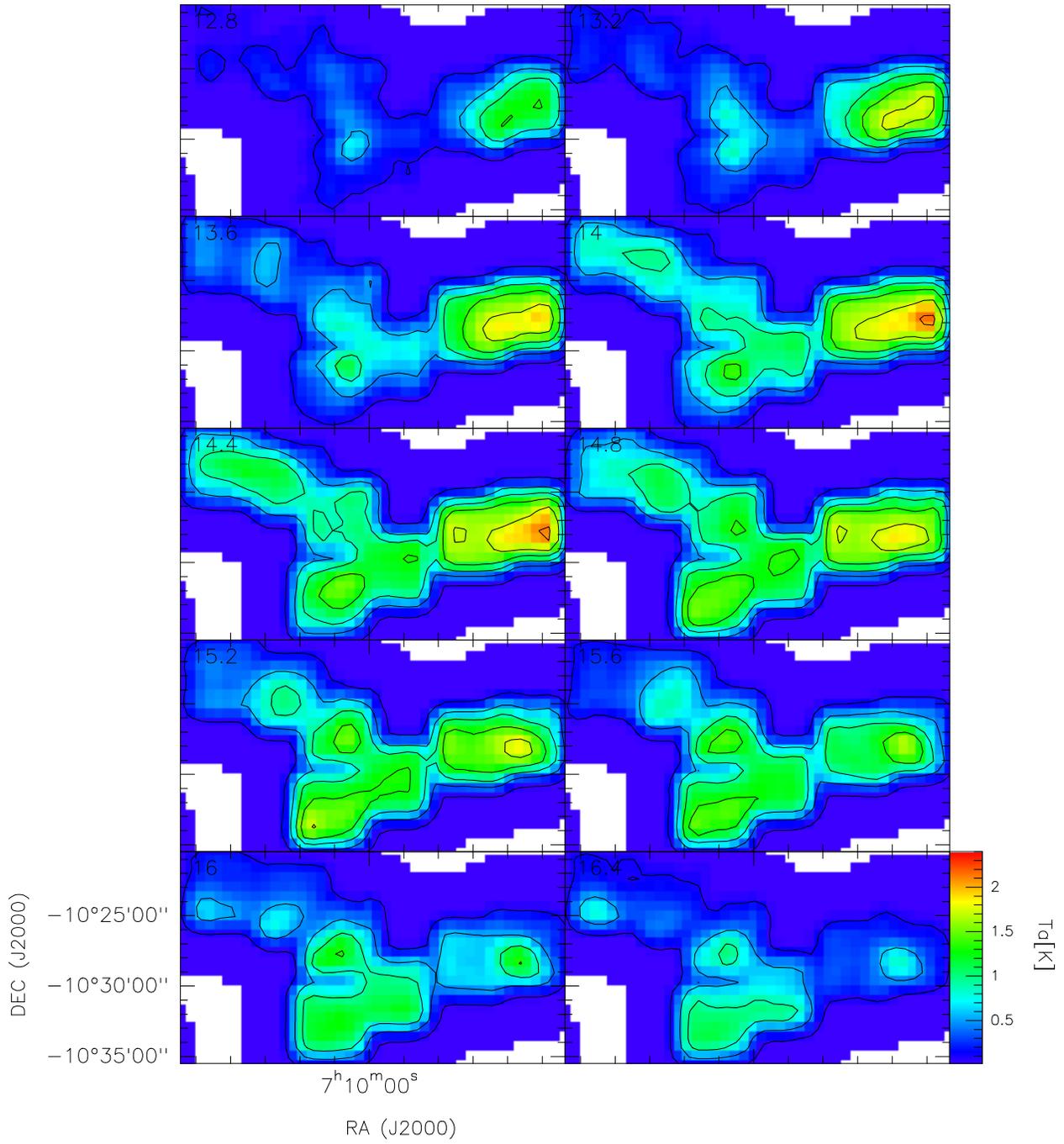}
\caption{
Same as Fig.\ref{fig:chmap13CO}, after convolving the map with a 80\,arcsec beam
and re-gridding the map on a 40\,arcsec step.
}
\label{fig:chmap13COconv}
\end{figure*}

%
%
\begin{figure*}
\centering
\hspace*{-0.5cm}
\includegraphics[width=18.0cm,angle=270]{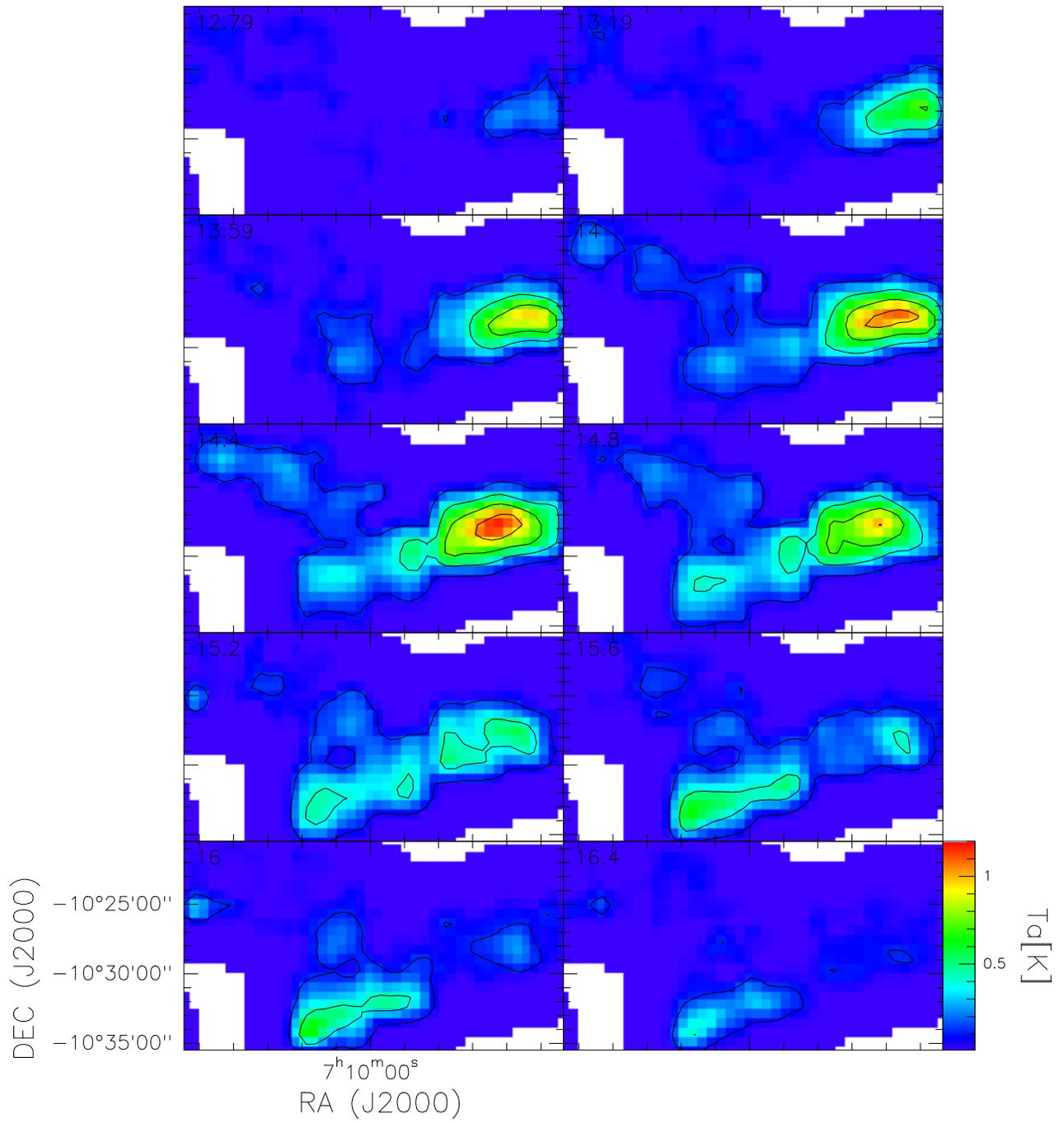}       
\caption{
Same as Fig.\ref{fig:chmap13COconv} for C$^{18}$O$(1-0)$.
}
\label{fig:chmapC18Oconv}
\end{figure*}

%
%
\begin{figure*}
\centering
\hspace*{-0.5cm}
\includegraphics[width=18.0cm,angle=270]{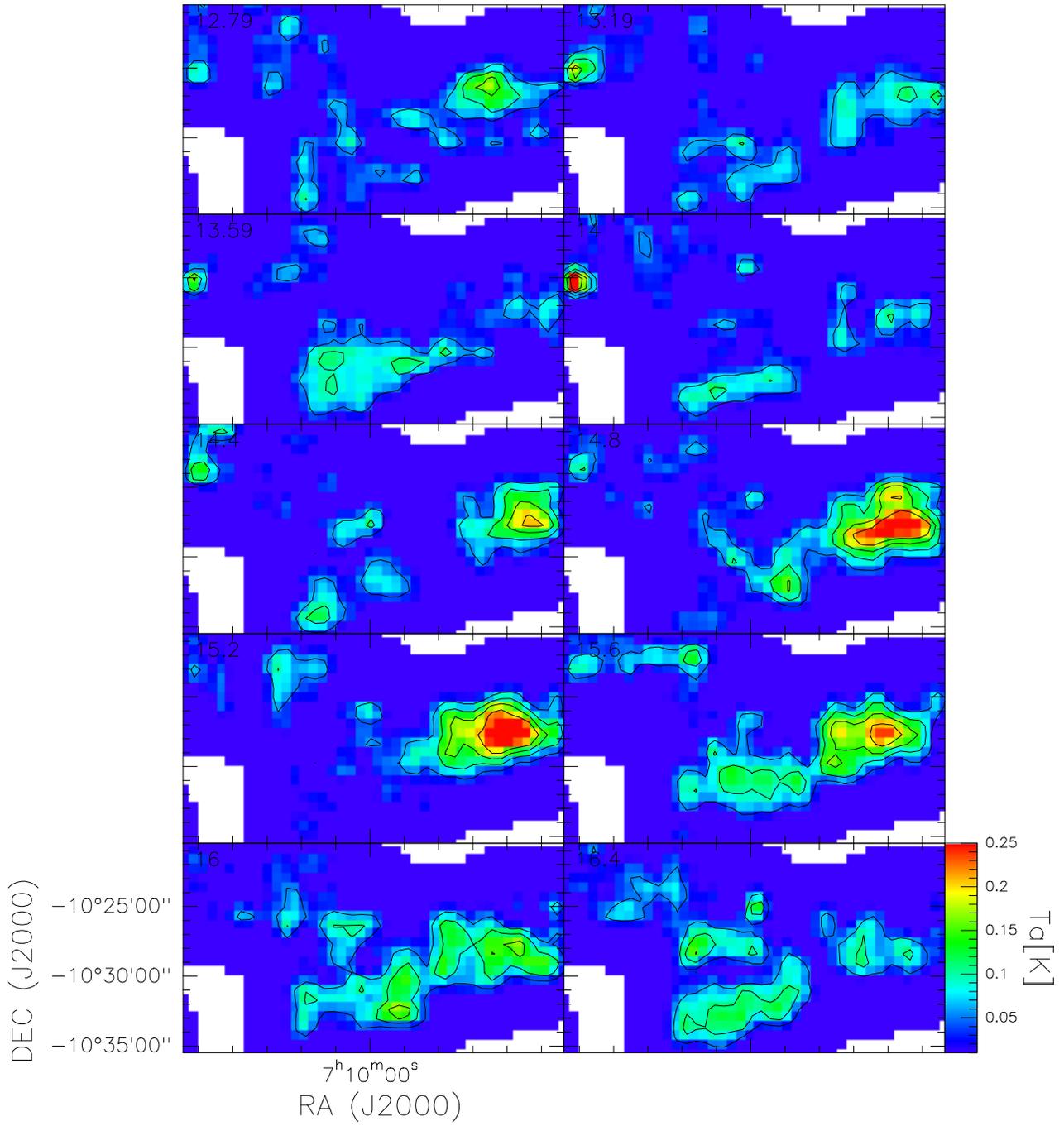}        
\caption{
Same as Fig.\ref{fig:chmap13COconv} for C$^{17}$O$(1-0)$.
}
\label{fig:chmapC17Oconv}
\end{figure*}

\end{document}